\documentclass[aps,10pt,prb,floatfix,citeautoscript,superscriptaddress,twocolumn,nofootinbib,longbibliography]{revtex4-2}

\usepackage{graphicx}
\usepackage{xcolor}
\usepackage{mathtools}
\usepackage{amsmath}
\usepackage{bbold}
\usepackage{hyperref}

\usepackage{relsize}
\relscale{1.5}

\newcommand{\dou}{\partial}

\newcommand{\bra}[1]{{{\langle #1 |}}}
\newcommand{\ket}[1]{{| #1 \rangle}}
\newcommand{\braket}[2]{\langle #1 | #2 \rangle}

\newcommand{\hx}{\hat{x}}
\newcommand{\hy}{\hat{y}}

\newcommand{\rr}{\boldsymbol{r}}
\newcommand{\qq}{\boldsymbol{q}}
\newcommand{\kk}{\boldsymbol{k}}

\renewcommand{\AA}{\boldsymbol{A}}

\newcommand{\yhat}{\hat{\boldsymbol{y}}}

\newcommand{\zhat}{\hat{\boldsymbol{z}}}

\newcommand{\term}[1]{\left( #1 \right)}
\newcommand{\abs}[1]{\left| #1 \right|}
\newcommand{\comm}[1]{\left[ #1 \right]}
\newcommand{\curly}[1]{\left\{ #1 \right\}}
\newcommand{\expect}[1]{\left\langle #1 \right\rangle}
\renewcommand{\d}{\partial}
\newcommand{\half}{\frac{1}{2}}

\newcommand{\eqnref}[1]{Eq.~(\ref{#1})}
\newcommand{\figref}[1]{Fig.~\ref{#1}}
\newcommand{\sfigref}[2]{Fig.~\hyperref[#1]{\ref{#1}#2}}

\newcommand{\secref}[1]{Sec.~\ref{#1}}
\newcommand{\appref}[1]{Appendix~\ref{#1}}

\usepackage{xcolor}
\definecolor{mydarkred}{RGB}{100,0,0}
\definecolor{mydarkblue}{RGB}{0,0,100}
\definecolor{mydarkviolet}{RGB}{150,30,60}
\definecolor{mydarkgreen}{RGB}{30,150,30}

\newcommand{\ix}{\tilde{x}}
\newcommand{\iy}{\tilde{y}}
\newcommand{\irr}{\tilde{\rr}}

\newcommand{\Caltech}{Department of Physics and Institute of Quantum Information and Matter, California Institute of Technology, Pasadena, CA 91125, USA.}

\newcommand{\mytitle}{Non-Fermi liquid and Weyl superconductivity from the weakly interacting 3D electron gas at high magnetic fields}

\begin{document}

\title{\mytitle}

\author{Nandagopal Manoj}\email{nmanoj@caltech.edu}\affiliation{\Caltech}
\author{Valerio Peri}\email{vperi@caltech.edu}\altaffiliation{Present address: Hexagon Innovation Hub GmbH, Heinrich-Wild-Strasse, 9435 Heerbrugg, Switzerland}\affiliation{\Caltech}
\author{Jason Alicea}\email{aliceaj@caltech.edu}\affiliation{\Caltech}

\date{\today{}}
\begin{abstract} 
Three-dimensional electron gases in strong magnetic fields host partially flat bands that disperse along the field direction yet exhibit Landau-level quantization in the transverse dimensions.  
Early work established that for spin-polarized electrons confined to the lowest Landau level band, repulsion triggers a charge density wave (CDW) in which electrons `self-layer’ into integer quantum Hall states, while attraction generates a non-Fermi liquid (rather than a superconductor).  
We revisit this problem with physically motivated deformations---including generalized local interactions, higher Landau level bands, restoration of spin, and explicit breaking of spatial symmetries---paying particular attention to the competition between CDWs and superconductivity.  
Our main findings are: 
(1) Generic local interactions can stabilize a nematic CDW in which integer quantum Hall layers spontaneously `tilt', yielding unconventional Hall response.  
(2) We numerically establish that the non-Fermi liquid appears stable to perturbations that preserve effective dipole conservation symmetries that emerge within a Landau level band. 
(3) Upon explicitly breaking translation symmetry, attraction catalyzes a novel layered superconductor that hosts Weyl nodes, superconducts within each layer, and insulates transverse to the layers.  
These results expand the rich phenomenology of interacting bulk electrons in the high-field regime and potentially inform the design of field-resistant superconductivity in low-carrier-density materials.

\end{abstract}


\maketitle

\tableofcontents

\section{Introduction}

In one-dimensional electron systems, interactions non-perturbatively modify low-energy properties and produce a non-Fermi liquid phase dubbed a Tomonaga-Luttinger liquid~\cite{Haldane1981}.  Two-dimensional systems can also experience non-perturbative interaction effects via application of a magnetic field that quenches the kinetic energy, leading to the quantum Hall effect in its myriad incarnations~\cite{GirvinYang2019}.  
These usually disparate phenomena converge in a three-dimensional electron gas subjected to a strong magnetic field such that the cyclotron energy becomes comparable to the Fermi energy~\cite{Halperin1987}.  There, the band structure deforms into a small number of partially occupied Landau level bands, which disperse with momentum along the field direction but remain highly degenerate due to quenched kinetic energy in the transverse directions.  The fate of the electron gas in this high-field regime upon turning on interactions poses a fundamental problem in quantum matter with a long history.

Partially quenched kinetic energy by the field produces a Fermi surface that is completely flat (see, e.g., Fig.~\ref{fig:Flat_FS}), consequently supporting a continuous family of density wave and Cooper pair nesting conditions. Resolving the competition between the density wave and superconducting tendencies presents a rich and subtle challenge. 
The earliest theoretical work studying interaction effects of bulk electrons in the high-magnetic-field regime dates back to Celli and Mermin~\cite{CelliMermin1965} in 1965. 
Within a Hartree-Fock treatment, these authors found that the ground state is unstable to the formation of a spin density wave with wavevector parallel to the magnetic field.
Other candidate orders such as charge density waves, valley density waves, exciton condensates, and Wigner crystals naturally emerge due to the nesting conditions, and were proposed to be relevant to various systems using a combination of mean-field, diagrammatic, and numerical techniques~\cite{Abrikosov1970,KleppmannElliot1975,Fukuyama1978,Kuramoto1978,SchlottmannGerhardts1979,Gerharsts1980,YoshiokaFukuyama1981,HeinonenAlJishi1986,MacDonaldBryant1987}. 
Crucially, it was noted by Halperin~\cite{Halperin1987} and later analyzed by Yakovenko~\cite{Yakovenko1991} that such density wave states %are topological and 
admit a quantized Hall response tied to the nesting wavevector.  In modern parlance, they comprise three-dimensional weak topological phases enabled by spontaneous symmetry breaking: Each `layer' of the density wave is smoothly connected to a 2D integer quantum Hall state, the collection of which yields a chiral `surface sheath' of gapless edge modes~\cite{BalentsFisher1996}. 

Meanwhile, the ground state of the electron gas with attractive interactions at high magnetic fields received relatively less attention. The enticing prospect of a superconducting instability~\cite{Rasolt1987,TesanovicRasoltXing1989,TesanovicRasoltXing1991,RasoltTesanovicRMP1992} was studied using mean-field and ladder approximations, which predict the formation of a spin-triplet superconductor in the high-field limit.  
Yakovenko~\cite{Yakovenko1993} employed a more controlled renormalization group (RG) approach---enabled by the residual dispersion and valid generally at weak interaction strengths---to study instabilities in the lowest Landau level band with contact interactions.  He made the crucial observation that, unlike the repulsive scenario, the mean-field and ladder approximations for the superconducting instability break down due to competition from the density wave channel. Whereas repulsive interactions catalyzed a density wave instability consistent with mean-field theory expectations, attraction generated a non-Fermi liquid (NFL) phase rather than a superconductor.

Several recent developments motivate revisiting this problem from a modern lens. 
With the advent of twisted moir\'e materials~\cite{MoireReview2021}, there is renewed interest in the many-body problem within almost flat bands, particularly the origin of superconductivity~\cite{PoZouSenthilVishwanath2018,BalentsReview2020}.  The three-dimensional electron gas enables a theoretically controlled RG analysis of correlations and pairing in partially flat-band systems, potentially offering useful insights into related settings.  
A growing number of bulk low-carrier-density materials (e.g., Weyl semimetals~\cite{ArmitageMeleVishwanathRMP2018}) reach the high-field limit at accessible magnetic field strengths, thereby expanding the platforms relevant for such studies. 
And on the theoretical side, there is a sharpened understanding~\cite{StahlLakeNandkishore2022,KapustinSpodyneiko2022,Spodyneiko2023} of the consequences of unconventional symmetries, such as magnetic translation symmetry and its relation to higher-moment conservation within the lowest Landau level, which one might hope to import over to the high-field three-dimensional electron gas. 

The main goal of this paper is to extend Yakovenko's functional RG techniques to assess the competition between density wave and Cooper channels in a more general class of interacting problems, in particular to explore the stability of the NFL state and conditions required for the appearance of high-field superconductivity.  We allow for various physically motivated deformations including generalized local interactions (which we show define an infinite family of marginal couplings), projection to higher Landau level bands, inclusion of spin, and explicit breaking of translation symmetry transverse to the field direction via a periodic potential.  A projected Landau level exhibits a pair of effective dipole conservation symmetries~\cite{Read1998,Spodyneiko2023}---one for each dimension transverse to the field.  Among the deformations we consider, the periodic potential is special in that it violates dipole conservation along one direction; all others preserve both dipole symmetries.   

Our analysis reveals that tuning the interaction range or considering higher Landau level bands can stabilize a \emph{nematic} topological charge density wave (CDW) composed of integer quantum Hall layers that spontaneously `tilt', producing an unconventional Hall response.  Previous work has captured such instabilities~\cite{Gerharsts1980}, though we add a fresh perspective on their origin and phenomenology.  More importantly, we provide numerical evidence that the NFL realizes a stable phase when the effective dipole conservation symmetries are fully preserved. In the case with spin included, we additionally uncover an intriguing quantitative connection between the stability requirements for the NFL and a strictly one-dimensional spinful Luttinger liquid.  

Adding a periodic potential bludgeons the density wave channel and allows Cooper processes to dominate \cite{Yakovenko1993}, unambiguously yielding superconductivity at weak coupling.  Here one can profitably view the periodic potential as nucleating superconducting `islands' oriented along the field direction; see Fig.~\ref{fig:Layered_SC}. This phenomenology is closely related to that of layered superconductors in a strong in-plane field~\cite{LebedYamaji1998}, and Ref.~\onlinecite{LeoLeon} studied a similar geometry using Ginzburg-Landau theory.  We show that our microscopic realization harbors several remarkable properties:  
The quasiparticle spectrum hosts gapless Weyl points.  Residual dipole conservation symmetry preserved by the periodic potential yields an unconventional coordinated inter-island Josephson coupling that precludes bulk phase stiffness transverse to the islands; that is, the system superconducts within each island but behaves like a `Bose-Einstein insulator'~\cite{LakeHermeleSenthil2022} in the orthogonal direction.  Finally, sample boundaries explicitly break even the residual dipole symmetry preserved by the periodic potential and can thus allow for transverse supercurrent confined to the surface.  

It remains unclear whether the NFL phase re-emerges beyond some critical interaction strength when a periodic potential is present, though we speculate that dipole conservation symmetries may provide the key principle underlying its stability.

\section{Continuum model}
\label{sec:model}

We first define the continuum model that underlies our analysis and that we will build on in later sections. Consider an interacting three-dimensional electron gas for spinless fermions $\Psi$ in a magnetic field ${\bf B} = B \zhat$, governed by a Hamiltonian $H = H_{\rm KE} + H_{\rm int}$ that respectively encodes kinetic energy and interactions. The kinetic part explicitly reads
\begin{equation}
    H_{\rm KE} = \int d^3\rr\, \Psi^\dagger(\rr) \comm{\frac{\term{-i\hbar\nabla - e \AA}^2}{2m} - \mu} \Psi(\rr),
\end{equation}
with $\AA = B x \yhat$ the vector potential expressed in Landau gauge, $m$ the electron mass, and $\mu$ the chemical potential. 
We specialize to the strong magnetic field---i.e., ``ultra-quantum''---limit such that the cyclotron frequency is the largest energy scale in the problem.  All the conduction electrons are then spin-polarized and reside in the lowest Landau level (LLL) band.

\begin{figure}
    \centering
    \includegraphics[width=1.0\linewidth]{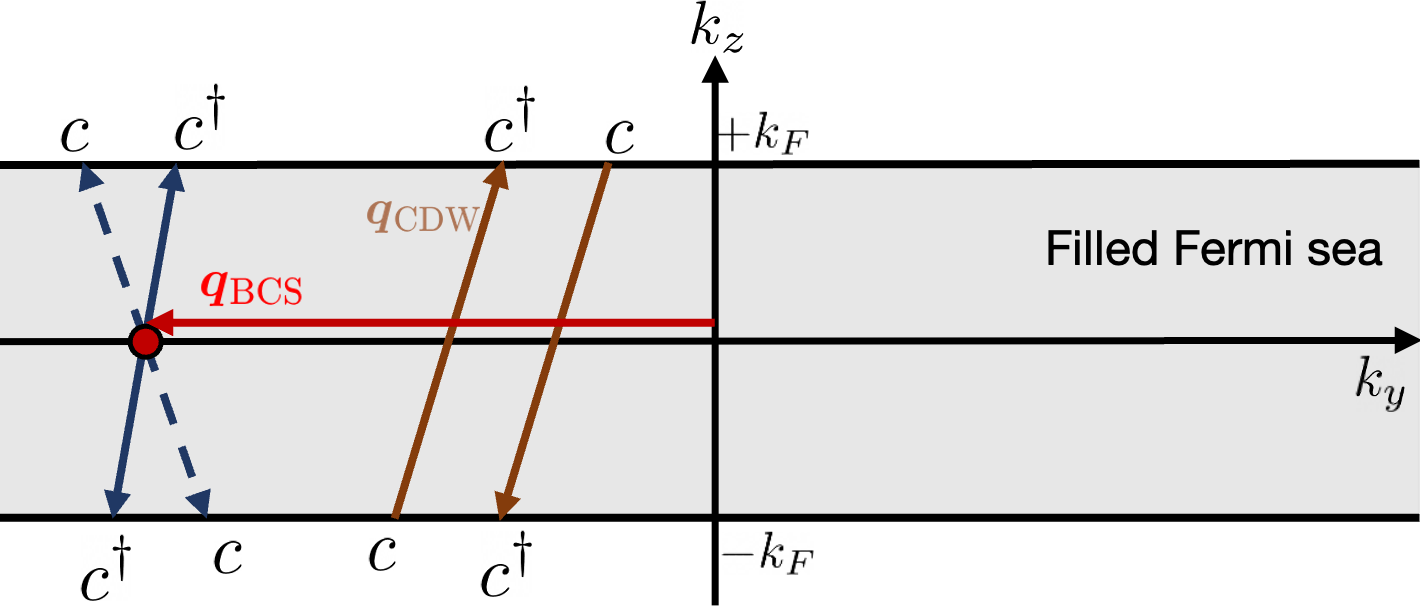}
    \caption{Flat Fermi surface of the non-interacting three-dimensional electron gas in a strong magnetic field. The blue arrows indicate BCS-type superconducting instabilities and the orange arrows indicate charge density wave instabilities.}
    \label{fig:Flat_FS}
\end{figure}

Given our gauge choice, $H_{\rm KE}$ yields LLL states labeled by $y$-direction momentum $k_y$ and localized around position 
\begin{equation}
  x = k_y l_B^2    
\end{equation} 
($l_B$ denotes the magnetic length). We define associated creation operators 
$\psi^\dagger_{k_y}(z_0)$ that, when acting on the electron vacuum, create a one-particle state with wavefunction
\begin{equation}
    \bra{\rr} \psi^\dagger_{k_y}(z_0) \ket{0} = \frac{1}{(\pi l_B^2)^{1/4}} e^{\comm{{i k_y y - (x- k_y l_B^2)^2/2}} } \delta(z-z_0) .
\end{equation}
Henceforth, we adopt units where $l_B = \hbar = 1$, occasionally restoring factors explicitly to make certain length or momentum scales manifest. Fourier transforming to momentum-space in the $z$ direction yields the LLL projected non-interacting Hamiltonian
\begin{equation}
    H_{\rm KE}^{\rm LLL} = \int_{k_y,k_z} \psi^\dagger_{k_y,k_z}  \left(\frac{k_z^2}{2m} - \mu \right) \psi_{k_y,k_z}.
\end{equation}
The conserved momentum quantum numbers $\kk \equiv (k_y,k_z)$ uniquely specify a single-particle state.  In terms of this two-dimensional momentum space, the non-interacting band structure is quasi-one-dimensional with a flat Fermi surface at the Fermi momenta $\pm k_F$; see \figref{fig:Flat_FS}.  Low-energy physics can be captured by linearizing the dispersion near the Fermi surface and defining slowly varying right- and left-moving fields $\psi_R, \psi_L$ via $\psi_{k_y}(z_0) \sim e^{i k_F z_0} \psi_{R,k_y}(z_0) + e^{-i k_F z_0} \psi_{L,k_y}(z_0)$.  
The corresponding non-interacting, low-energy Euclidean action is given by 
\begin{align}
    S_0 &= \int_{k_y,q_z} d\tau \, \big{[}\psi^\dagger_{R,k_y,q_z} (\dou_\tau - v_F q_z)   \psi_{R,k_y,q_z} \nonumber \\
    &\quad + \psi^\dagger_{L,k_y,q_z} (\dou_\tau + v_F q_z)  \psi_{L, k_y,q_z}\big{]}
    \label{eqn:S0}
\end{align}
with $v_F = k_F / m$ the Fermi velocity and $q_z$ a momentum measured relative to the Fermi points.

To study the effects of generic weak interactions $H_{\rm int}$ on this partially flat-band problem, we impose the following symmetries:
\begin{enumerate}
    \item $U(1)$ particle number conservation.
    \item Non-commuting magnetic translation and rotation symmetry in the $xy$ plane.
    \item Translation symmetry along $z$.
    \item Time reversal composed with reflection along $x$ ($\mathcal{M}_x\mathcal{T}$). 
    \item Reflections $\mathcal{M}_z$ along $z$. 
\end{enumerate}
We start from a general four-fermion interaction between right- and left-movers. In \appref{App:SymmetryAllowedInteractions}, we review consequences of the microscopic symmetries and show that they constrain the interaction to the form
\begin{multline}
    S_{\text{int}} = \int_{\{k_y,q_z\}} d\tau h({k_y}_1,{k_y}_2) \psi^\dagger_{R,k_y + {k_y}_1 / 2} \psi^\dagger_{L,k_y - {k_y}_1 / 2} \\
    \times\psi_{L,k_y - {k_y}_2 / 2} \psi_{R,k_y + {k_y}_2 / 2} 
     \delta(\Sigma q_z).
     \label{eqn:Sint}
\end{multline}
For brevity $q_z$ momenta for each field have been suppressed, and $\Sigma q_z$ is understood to enforce standard momentum conservation along the $z$ direction.  (One could alternatively derive Eq.~\eqref{eqn:Sint} by starting from a manifestly symmetric microscopic interaction and then manually projecting to the LLL.)  
Since the couplings depend only on $k_y$'s, the above interaction is local (contact, with no derivatives) in both $\tau$ and $z$. This simplification is justified because if we treat $S_0$ in \eqnref{eqn:S0} as the non-interacting fixed point, higher-derivative terms have higher scaling dimension and are therefore irrelevant at low energy. However, because the kinetic energy is flat with respect to $k_y$, there is no RG justification for assuming that in-plane interactions are dominated by the contact term; hence we allowed a general symmetry-allowed coupling function $h(k_{y_1},k_{y_2})$.  

Beyond symmetry, locality further constrains $h(k_{y_1}, k_{y_2})$ to decay exponentially with $k_{y_1}$ and $k_{y_2}$.  This constraint arises because the wavefunction of a particle with definite $k_y$ is exponentially localized at $x = k_yl_B^2$; consequently, any local interaction must produce a coupling function with exponential decay in the momentum transfer and relative momentum.

Before we begin the formal RG analysis, it is useful to gain some intuition about the competing potential instabilities, which manifest as marginal interactions at tree level. 
Namely, for arbitrary $q$, the Fermi surface has a nesting vector given by $\qq_{\text{CDW}} = q \yhat + 2 k_F \zhat$ promoting CDW order. We also have, for arbitrary $q$, a superconducting (BCS) type nesting of the Fermi surface such that the putative Cooper pairs condense with total momentum $2\qq_{\text{BCS}} = q\yhat$. See arrows in \figref{fig:Flat_FS} for an illustration of the CDW- and BCS-promoting interactions.

The intuition for the interacting theory is thus similar to the RG for the one-dimensional Fermi gas (Luttinger liquid) in that these potential BCS and CDW instabilities compete with each other.  The key difference is that there are now an extensive number of coupling constants to keep track of---corresponding to the extensive number of putative orders due to arbitrariness in $q$. In the one-dimensional interacting Fermi gas, it is known that the BCS and CDW diagrams cancel out at all loop orders in the renormalization group, giving rise to the spinless Luttinger liquid phase. This exact cancellation does not happen in the present three-dimensional problem: the coupling function $h$ evolves according to a nontrivial flow equation that one must integrate numerically. The RG flow for the spinless three-dimensional LLL interacting electron problem described next has been calculated using functional renormalization group techniques by Yakovenko in Ref.~\onlinecite{Yakovenko1993}, which we closely follow in the first part of this work.

\section{Functional renormalization group analysis}
\label{sec:FRG}

We begin this section by reviewing the work of Ref.~\onlinecite{Yakovenko1993}, reinterpreting it as Polchinski-Shankar Fermi surface RG~\cite{Polchinski1992,Shankar1994}. We describe how a general marginal coupling evolves under perturbative RG flow using a one-loop calculation, allowing us to perform an unbiased analysis of the properties of this state at long distances and low energies. There are three possibilities under RG: the coupling flows to zero (irrelevant), remains finite (marginal), or flows to strong-coupling (relevant). In the latter case, we stop running the RG once the coupling becomes 
large, where the flow of the coupling function becomes uncontrolled in the numerics. In practice, we calculate the initial norm of the coupling function $\int d^2\rr f(\rr)^2$ and stop running the RG when the norm becomes $\sim 1.1$-$1.2$ times the initial value, as this criterion suggests that the system is flowing to a strongly coupled phase. 

\subsection{Gaussian fixed point} 
Equations \eqref{eqn:S0} and \eqref{eqn:Sint} specify the full action including bare interactions whose impact we wish to assess. 
Performing Wilsonian RG to the infrared, we integrate out modes with $\Lambda / s < \abs{q_z} < \Lambda$, where $\Lambda$ is a momentum cutoff for the $z$ direction and $s>1$.  At the Gaussian fixed point with $h(k_{y_1},k_{y_2}) = 0$, this operation is trivial since the `fast' and `slow' modes  decouple. Upon rescaling to restore the original cutoff, we perform the usual scaling of momentum and frequency, except that the $k_y$ momenta are not rescaled, due to the flat dispersion along that direction. These transformations give the scaling dimension of various objects at the Gaussian fixed point.

At tree level, the interaction term of Eq.~\eqref{eqn:Sint} has scaling dimension 2 
as the momentum conservation constraint can be factorized into separate conservation laws for the momentum parallel and perpendicular to the Fermi surface, and the perpendicular component $\delta(\Sigma q_z)$ contributes scaling dimension $-1$.   (Note that scaling dimension 2 is marginal in this problem because we scale only two of the spacetime directions.) A similar decomposition into two perpendicular axes cannot be done in curved Fermi surfaces, which renders most interactions irrelevant~\cite{Polchinski1992}. Another way to understand this result is that Pauli blocking is not effective when one has a flat Fermi surface, as there are many scattering processes near the Fermi surface that conserve both energy and momentum. Therefore, every interaction of the form in Eq.~\eqref{eqn:Sint} is marginal at tree level because of the {flat Fermi surface}, and one must systematically account for how this coupling function renormalizes at one loop.  

\subsection{Functional RG flow at one loop}
 At one loop, there are two diagrams that contribute to renormalizing the coupling function $h(k_{y_1},k_{y_2})$ in the interaction term~\cite{Yakovenko1993}. One diagram corresponds to coherent propagation of a right-moving electron, left-moving hole pair (CDW term) and the other corresponds to coherent propagation of an electron-electron pair (BCS term). The RG flow due to the two processes follows upon integrating out `fast' internal legs and can be concisely written as 
\begin{equation}\label{eqn:RGflow1}
    \frac{d h({k_y}_1,{k_y}_2)}{d\xi} = \frac{\comm{(h \cdot h)({k_y}_1,{k_y}_2) - (h \ast h)({k_y}_1,{k_y}_2)}}{2 \pi v_F},
\end{equation}
where $\xi = \log(s)$ denotes the RG time and we use the functional product notation~\cite{AliceaBalents2009}
\begin{widetext}
\begin{equation}
\begin{aligned}
    (h \cdot h)({k_y}_1,{k_y}_2) &\equiv \sqrt{\frac{\pi}{2}} \int dk'_y \; h\term{k'_y + \frac{{k_y}_1+{k_y}_2}{2}, -k'_y + \frac{{k_y}_1+{k_y}_2}{2}}  h\term{-k'_y + {k_y}_1 , k'_y + {k_y}_2}, \\
    (h \ast h)({k_y}_1,{k_y}_2) &\equiv  \sqrt{\frac{\pi}{2}} \int dk'_y \;  h\term{{k_y}_1,k'_y} h\term{k'_y,{k_y}_2}.
\end{aligned}
\label{eqn:shorthand}
\end{equation}
\end{widetext}
We now repackage the coupling function in terms of alternative representations $\lambda,\gamma,\Gamma$, related by linear transformations:
\begin{align}
    \lambda({k_y}_1,{k_y}_2) &\equiv h\term{{k_y}_1 + {k_y}_2,{k_y}_2 - {k_y}_1} , \nonumber \\
    \gamma(\ix,\iy) &\equiv \frac{1}{\sqrt{2\pi}} \int dk_y e^{i k_y \iy} \lambda(\ix, k_y), \nonumber \\
    \Gamma(k_x,k_y) &\equiv \frac{1}{2\pi} \int d\ix d\iy  e^{-i (k_x \ix + k_y \iy)} \gamma(\ix,\iy).
    \label{eqn:gamma_rep}
\end{align}
In particular, $\gamma(\irr)$ is rotationally invariant in parameter space due to the Hamiltonian's $U(1)$ rotation symmetry in the $xy$ plane, as we show in \appref{App:SymmetryAllowedInteractions}. The functional products $\cdot$ and $\ast$ will have different forms under these representations, and in terms of $\gamma$ we obtain~\cite{Yakovenko1993}
\begin{equation}
    \frac{d \gamma(\irr)}{d\xi} = \frac{1}{2 \pi v_F}\int d^2\irr' \gamma(\irr')\gamma(\irr-\irr')\term{1- e^{i \irr\wedge \irr'}},
\end{equation}
where $\irr \equiv (\ix,\iy)$ and $\wedge$ denotes the wedge product $\irr\wedge \irr' \equiv \ix\iy' - \iy\ix'$.

Some aspects of this RG flow can be understood analytically, but one needs to perform numerical integration to understand the full flow around the Gaussian fixed point. Reference \onlinecite{Yakovenko1993} considered the special case of contact density-density interactions between the right- and left-moving fermions fields and found the following:
\begin{enumerate}
    \item \textit{Repulsive interactions}: The CDW instability overpowers the BCS instability and the system flows to strong coupling. A mean-field analysis of the renormalized interaction suggests that the system spontaneously layers along $z$, i.e., it has a CDW nesting instability with momentum $2 k_F \zhat$. In the limit of large CDW order parameter (i.e., decoupled layers), each layer can be thought of as a two-dimensional $\nu = 1$ integer quantum Hall state. In the rest of the article, we call this phase a \emph{normal} topological CDW state. 
    \item \textit{Attractive interactions}: Neither instability diverges and the system enters a gapless NFL state\footnote{Reference~\onlinecite{Yakovenko1993} referred to this state as a marginal Fermi liquid.  Typically, in a marginal Fermi liquid the quasiparticle residue vanishes logarithmically at the Fermi surface as you take $\omega \to 0$. In this system, however, it vanishes faster (exponential of polylog).}. 
    The origin of the NFL and its stability to perturbations remain poorly understood. 
\end{enumerate}

\section{Beyond contact interactions}
Often contact interactions yield the most relevant terms, with those containing derivatives suppressed because of the positive scaling dimension of spatial and temporal derivatives under Wilsonian RG.  As noted earlier, this intuition (partially) fails in the present problem because of the flat dispersion of the LLL: transverse derivative terms in the unprojected interaction are not punished under RG. Is the NFL stable to such perturbations?  
And can these terms produce new phases beyond those accessible from contact interactions?  

\begin{figure}
    \centering
    \includegraphics[width=0.9\linewidth]{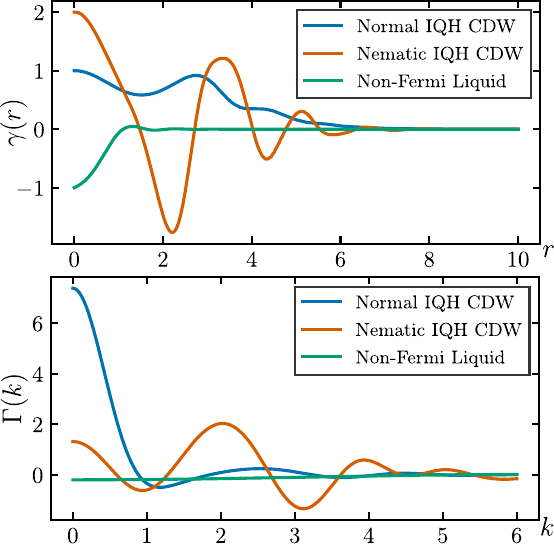}
    \caption{Typical renormalized coupling functions for the three phases discussed in this section, evaluated using the bare coupling (\ref{eqn:g0g1}) with $(g_0,g_1) = (1,0)$ for the blue curve, $(0,1)$ for the orange curve, and $(-1,0)$ for the green curve.  The bottom panel is the two-dimensional Fourier transform (Hankel transform) of the top panel.}
    \label{fig:RCFs}
\end{figure}

The simplest microscopic (i.e., not yet LLL-projected) interaction containing transverse derivatives and satisfying all the symmetries is 
\begin{equation}
    H_\text{int} = g_1 \int d^3\rr \psi^\dagger_R(\rr)  [D_+\psi^\dagger_L(\rr)][ D_-\psi_L(\rr)] \psi_R (\rr),
    \label{eq:g1}
\end{equation}
where $\rr$ is a three-dimensional spatial index (not to be confused with $\tilde \rr$) and $D_\pm$ denotes the covariant derivative $D_x \pm i D_y$. Note that $\psi^\dagger$ and $\psi$ carry opposite electric charge; therefore the covariant derivatives acting on them look different. Adding this term to the pure contact interaction with strength $g_0$ and projecting to the LLL yields the bare interaction
\begin{equation}
\label{eqn:g0g1}
    \gamma_0(\tilde \rr) =\comm{g_0 +  g_1(2-\tilde \rr^2)} e^{-\tilde \rr^2 / 2}.
\end{equation}
Upon calculating the RG flow numerically in the $(g_0,g_1)$ space, we find three distinct behaviors of the renormalized coupling function illustrated for typical cases in Fig.~\ref{fig:RCFs}.  The resulting phase diagram in Fig.~\ref{fig:PD1} displays two noteworthy features: First, with attractive $g_0<0$, the NFL phase appears stable over a finite window of $g_1$---\emph{and hence does not require contact interactions}. This stability provides strong evidence, further built upon in later sections, that the NFL constitutes a gapless phase as opposed to a fine-tuned point in parameter space. 

Second, $g_1$ enables a new phase that we will identify as a nematic cousin of the normal topological CDW. In this new phase, the renormalized coupling function $\gamma_{\text{ren}}(\rr)$ is isotropic and appears as in the orange curves of \figref{fig:RCFs}.
Its two-dimensional Fourier transform $\Gamma_{\text{ren}}(\kk) \propto \int_{\rr} e^{i\kk\cdot\rr} \gamma_{\text{ren}}(\rr)$ strongly peaks along a circle at $\abs{\kk} = k_c$ with $k_c$ a non-universal wavevector.  One can approximate $k_c$ by neglecting the BCS diagram in the flow equation, which is justified as the numerical flow shows that the CDW diagram dominates.  Within this approximation, $k_c$ is given by the 
magnitude of $\kk$ where the bare Fourier transform $\Gamma_0(\kk)$ is maximized, yielding
\begin{equation}
    k_c \approx \sqrt{\frac{2 g_1 - g_0}{g_1}}.
\end{equation}
One can also check that the $k_c$ in the numerics quantitatively matches the above approximate result when $k_c$ is small.
In what follows we describe a self-consistent mean-field solution of the renormalized coupling function.

\subsection{Self-consistent mean field theory}
The renormalized interaction is given by
\begin{widetext}
\begin{align}
    H_{\text{int,ren}} = \int_{\{k_y,q_z\}}  h_{\text{ren}}({k_y}_1,{k_y}_2) \psi^\dagger_{R,k_y + {k_y}_1 / 2} \psi^\dagger_{L,k_y - {k_y}_1 / 2} 
    \psi_{L,k_y - {k_y}_2 / 2} \psi_{R,k_y + {k_y}_2 / 2} 
     \delta(\Sigma q_z),
     \label{eqn:renint1}
\end{align}
\end{widetext}
where $q_z$ is integrated up to the renormalized cutoff and $h_\text{ren}$ is related to the peaked $\Gamma_\text{ren}$ by a linear transformation through \eqnref{eqn:gamma_rep}. To make this problem analytically tractable, we assume that the ground state spontaneously picks a direction in the $\Gamma(\kk)$ representation, and then minimize the energy for this interaction. It is possible that the system picks a different pattern of symmetry breaking, such as forming a multi-component CDW~(see Ref.~\onlinecite{Halperin1987} and references therein),
but we restrict ourselves to the simplest scenario here. Our emphasis is on the observation that even after restricting to local interactions and the lowest Landau level band, details of the interaction (which are affected by details such as the geometry of the electronic orbitals) play a decisive role in determining the nature of the ground state order. 

Assuming that the ground state spontaneously picks one direction, the relevant part of the effective interaction can be written as $\Gamma(\kk) = g \delta^2(\kk - \kk_c)$, where 
\begin{equation}
  \kk_c = k_c(\cos \theta, \sin \theta )
  \label{eq:kc}
\end{equation}
with arbitrary, but fixed, $\theta$. 

Plugging the corresponding $h({k_y}_1,{k_y}_2)$ into \eqnref{eqn:renint1}, we decouple the renormalized interaction in the CDW channel to obtain (up to a constant) 
\begin{equation}
    H_{\rm int,ren} \approx \frac{g}{(2\pi)^{3/2}} (O^\dagger \langle O\rangle + \langle O^\dagger \rangle O)
\end{equation}
with
\begin{align}
    O \equiv \int_{q_z,k_y}  e^{- i k_y k_c \cos\theta }\psi^\dagger_{L,k_y -\frac{k_c \sin\theta}{2}, q_z}  \psi_{R,k_y + \frac{k_c \sin\theta}{2}, q_z} 
\end{align}
the CDW order parameter.
Defining the rescaled order-parameter expectation value $\Delta \equiv g \expect{O} /(2\pi)^{3/2}$, we can solve the self-consistent gap equation at zero temperature and obtain 
\begin{equation}
    |\Delta| \sim \Lambda(\xi) e^{-\frac{1}{g(\xi) N(0)}} 
\end{equation}
where $\xi$ is the RG time, $\Lambda(\xi)$ and $g(\xi)$ are the renormalized UV cutoff and interaction strength respectively, and $N(0)$ is the density of states at the Fermi level. 

\subsection{Phenomenology of the nematic topological CDW}
\label{sec:nCDW}
Calculating the density profile of this state, we find
\begin{equation}
    \expect{\Psi^\dagger(\rr)\Psi(\rr)} \sim \cos(\kk \cdot \rr + \varphi), 
\end{equation}
where $\varphi$ is the phase of the CDW order parameter $\langle O\rangle$ and $\kk = (-\kk_c, 2 k_F)$ is the CDW wavevector (we refer the reader to \appref{app:tCDW} for more details on this calculation). The arbitrary choice of $\varphi$ by the mean-field ground state breaks one component of translation symmetry and has an associated linearly dispersing Goldstone mode. 

The arbitrary choice of $\theta$ in $\kk_c$---recall Eq.~\eqref{eq:kc}---spontaneously breaks $U(1)$ rotation symmetry in the $xy$ plane (nematicity), and has experimentally accessible consequences. One can intuitively view this state as adiabatically connected to a decoupled stack of $\nu = 1$ integer quantum Hall layers, each tilted away from the $(x,y)$ plane---i.e., it realizes a nematic topological CDW as claimed earlier. Since the tilting is specified by the wavevector $\kk_c + 2k_F\zhat$, we find $\sigma_{xy} = (e^2 / h)(k_F/\pi) = n e / B$ as expected classically from Galilean invariance. There is also an \emph{anomalous} component to the Hall conductivity \cite{Halperin1987}, which is a signature of spontaneous rotational symmetry breaking: Measuring the Hall conductivity along the plane perpendicular to $(\kk_c,0)$ yields $\sigma_{\hat{\boldsymbol{\theta}} z} =  (e^2 / h) (k_c/2\pi)$, with $k_c \sim l_B^{-1}$ (upon restoring the magnetic length) a non-universal value.

Although this state breaks more continuous symmetries than the normal IQH CDW state, we do not expect any more Nambu-Goldstone modes as the additional broken symmetry is rotational (around the $z$ axis) and does not commute with the broken translation symmetry generator. The absence of an additional Nambu-Goldstone mode follows from the nontrivial Watanabe-Brauner matrix~\cite{WatanabeBrauner2011,SSB2019}. 

This analysis shows that, even with short-ranged repulsive interactions, the IR physics is sensitive to the exact form of the interactions between the UV fermionic fields. The situation differs from more conventional field theories where the IR physics is usually captured using the lowest-order term in a derivative expansion of the four-fermion interaction term, as alluded to in the previous section. 

\begin{figure}
    \centering
    \includegraphics[width=0.85\columnwidth]{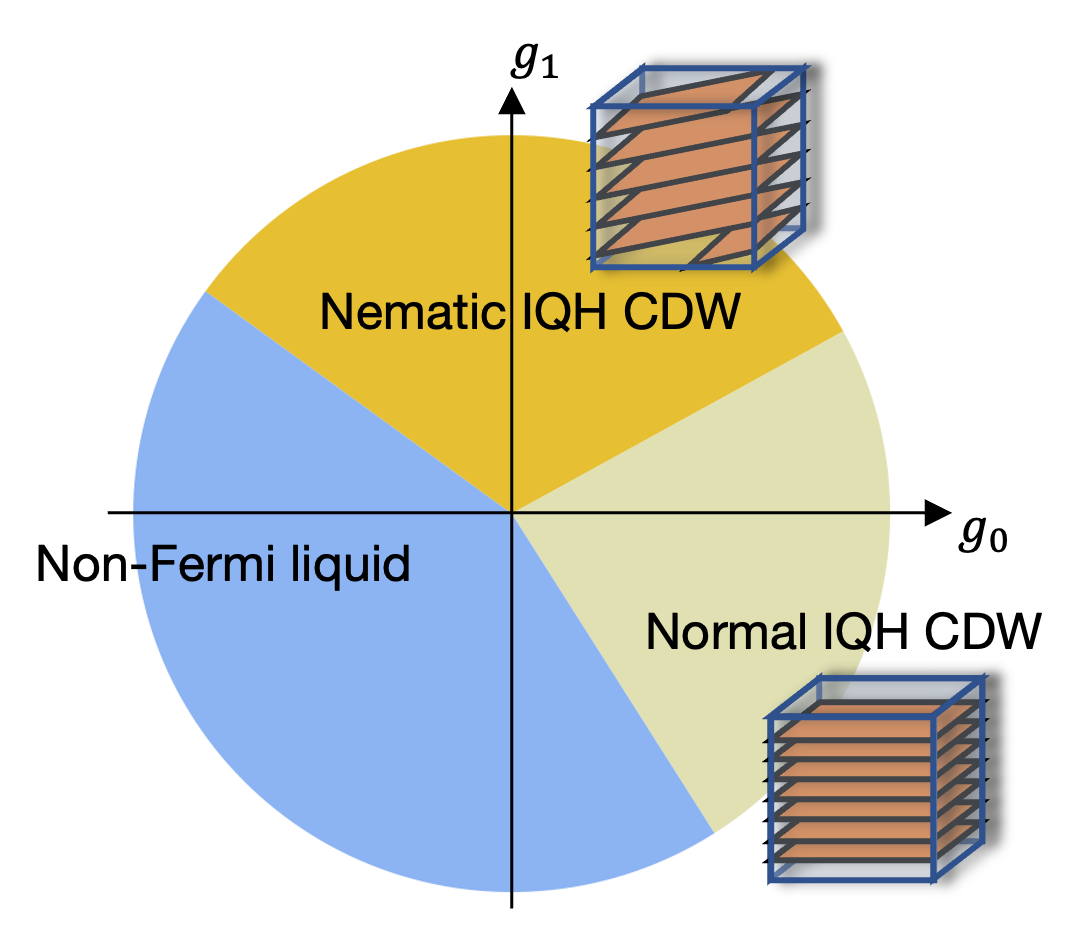}
    \caption{Phase diagram for the weakly interacting three-dimensional electron gas in the lowest Landau level determined via numerical integration of the flow equations. The horizontal axis represents the contact density-density interaction studied previously \cite{Yakovenko1993}, while the vertical axis represents a higher derivative interaction [\eqnref{eq:g1}] that enables a nematic integer quantum Hall charge density wave phase.  A qualitatively similar phase diagram emerges when projecting those interactions into the second Landau level instead of the lowest Landau level.}
    \label{fig:PD1}
\end{figure}

\section{Attractive Instabilities} \label{sec:attractiveinstabilities}
After finding in the previous section that the IR of the RG flow depends on details of the microscopic interaction, we pivot to a new puzzle: in our numerical investigation, we observe that the RG flow equations never diverge to an attractive instability. In the case of attractive density-density interactions between the right- and left-moving fermion fields, we instead find, as in Ref.~\onlinecite{Yakovenko1993}, that the system does not spontaneously break any symmetries and enters a NFL state.  This phase persisted even when deforming the interaction range via inclusion of nonzero $g_1$---the first hint that the NFL represents a stable phase.  
Surprisingly, we did not find superconducting instabilities that spontaneously break charge conservation symmetry, henceforth dubbed $U(1)_C$.
Next we briefly survey a few possible resolutions to the absence of superconductivity and then introduce several other variations of the problem to further explore the NFL state's stability and the propensity towards pairing. 

Reference~\onlinecite{Spodyneiko2023} discusses $U(1)_C$ spontaneous symmetry breaking in the context of the LLL in \emph{two} spatial dimensions. It is shown that, in the translation-invariant LLL, there is a nontrivial commutation relation linking the magnetic translation symmetries and the total charge, equivalent to the algebra of momentum and dipole operators in one dimension:
\begin{equation}
    \comm{P_x, P_y} = i l_B^{-2} Q,
    \label{eq:PPQ}
\end{equation}
where $P_i$ are the generators of magnetic translations and $Q$ is the total charge.  For a one-dimensional charge-$q$ particle with position operator $X$, momentum operator $P$, and dipole operator $d = q X$, we have $[d,P] = i q$.   Comparing to Eq.~\eqref{eq:PPQ}, we can identify $-l_B^2 P_y \equiv D_x$ as the dipole moment operator in the $x$ direction, and similarly identify $l_B^2 P_x \equiv D_y$ as the dipole moment along $y$. Consequently,  continuous translation symmetry in the LLL is \emph{equivalent to} dipole conservation symmetry. The conserved (gauge-invariant) momentum quantum number along $x$ is the same as the dipole moment along $y$ (up to factors), and vice versa. We recall this perspective later when we discuss explicit breaking of translation symmetry in \secref{sec:6breakingtranslation}. 

Equation~\eqref{eq:PPQ} implies that $U(1)_C$ spontaneous symmetry breaking cannot occur if one preserves magnetic translation symmetry, as one cannot preserve dipole moment without preserving charge.  Correspondingly, $U(1)_C$ plays a different role compared to other $U(1)$ symmetries in the problem (e.g., spontaneous translation symmetry breaking in a CDW state).  In the presence of this effective dipole conservation symmetry, Hohenberg-Mermin-Wagner-type theorems can be modified to argue that spontaneous breaking of $U(1)_C$ symmetry is forbidden even in $2+1$D~\cite{Spodyneiko2023}.
Unfortunately, such generalized Hohenberg-Mermin-Wagner theorems do not explain the lack of $U(1)_C$ spontaneous symmetry breaking in our problem, which takes place in three spatial dimensions. A heuristic argument goes as follows (we refer the reader to Refs.~\onlinecite{Spodyneiko2023,KapustinSpodyneiko2022,StahlLakeNandkishore2022} for detailed discussion): if we write down an effective field theory of the putative charged superfluid in the LLL~\cite{Moroz2018,Du2024,Manoj2025,GlodkowskiLLLSuperfluidCoset2026} in terms of a low-energy phase degree of freedom $\phi$, the action takes the form (up to non-universal prefactors)
 \begin{equation}
     S[\phi] = \int d^2\rr_{\perp} d^{D-2}\rr_{\parallel} d\tau \half\comm{(\partial_\tau \phi)^2 + (\nabla_\perp^2 \phi)^2 + (\nabla_\parallel \phi)^2},
 \end{equation}
 where the quadratic in-plane derivative term is forbidden by the dipole symmetry $\phi(\rr,\tau) \mapsto \phi(\rr,\tau) + \boldsymbol{\beta}\cdot \rr_\perp $, in turn enhancing fluctuations. Specifically, for $D = 2$, the phase fluctuations exhibit a logarithmic infrared divergence $\expect{\delta\phi^2} \sim \int_{\omega, \kk} G(\omega, \kk)$. Since this divergence is absent for $D=3$, the Hohenberg-Mermin-Wagner theorem does not preclude a superconducting instability in the three-dimensional problem of interest here.

 Our numerical simulations are only valid close to the non-interacting limit, where the analysis is controlled.  It is possible that a superconducting instability in the problem we have analyzed so far requires strong interactions.  Next we consider more modest alternatives and ask whether a \emph{weak-coupling} attractive instability can set in upon altering the problem in various physically motivated ways. We stress that all of the scenarios we consider for the remainder of this section fully preserve the effective dipole conservation symmetries highlighted above.

\subsection{Higher Landau level bands}

Perhaps projecting attractive real-space interactions into \emph{higher} Landau level bands can produce a weak-coupling instability that breaks $U(1)_C$?  Higher Landau level bands exhibit different single-particle wavefunctions compared to the LLL band---in turn yielding a distinct set of initial coupling functions even for the same microscopic parameters. We have explored the variation where $g_0$ (contact interaction) and $g_1$ (Eq.~\eqref{eq:g1}) are projected into the second Landau level band, and interactions with the LLL band are neglected for simplicity.  The corresponding bare coupling function reads
\begin{align}\label{eqn:g0g1_1LL}
    \gamma_0(\tilde \rr) &= \comm{g_0\term{1-\frac{\tilde r^2}{2}}^2 + \frac{g_1}{2}\term{1-\frac{\tilde r^2}{2}}\term{\tilde r^4 -8\tilde r^2 +8}} \nonumber \\
    &\qquad \qquad \times  e^{-\tilde \rr^2 /2}.
\end{align}
The phase diagram determined via numerical evaluation of the flow equations exhibits the same structure as Fig.~\ref{fig:PD1}, but with quantitative shifts in phase boundaries.  Evidently qualitatively similar phenomenology arises within the LLL and second Landau level bands---including an extended NFL phase and the conspicuous absence of superconducting instabilities even when the bare interactions are attractive. 

\subsection{Explicitly breaking rotation symmetry}
Returning to the LLL problem, next we explore stability of the NFL state to perturbations that explicitly break continuous rotation symmetry.  In the LLL, rotation symmetry is related to conservation of a second moment of charge, $\int d^2 \rr \; r^2 \rho(\rr)$ as the generator of magnetic rotations is given by $(\hat{x}^2 + \hat{y}^2)/2$, where $\hat{x},\hat{y}$ are the LLL-projected position (guiding center) operators. 
One might then wonder whether the NFL is stabilized by a lack of phase space for the electrons to scatter to, due to the higher-moment conservation laws
imposed by the magnetic translation and rotation symmetries. Such a stability criterion would be loosely analogous to that of a Fermi liquid, where momentum conservation and a curved Fermi surface
imply a lack of low-energy scattering processes. 

\begin{figure}
    \centering
    \includegraphics[width=0.95\columnwidth]{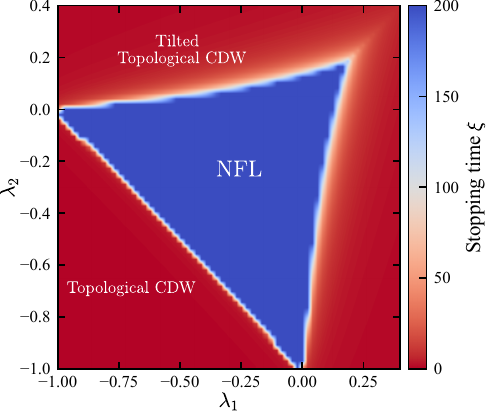}
    \caption{Stability of the non-Fermi liquid phase to rotation-symmetry-breaking perturbations encoded in the bare coupling function from \eqnref{eqn:anisotropic} when $\lambda_1 \neq \lambda_2$. Here, the ``tilted'' phase is not a nematic as the rotation symmetry is explicitly broken; it only spontaneously breaks the residual $\mathbb{Z}_2$ symmetry of $\pi$ rotations. }
    \label{fig:anisotropicPD}
\end{figure}

We explore this possibility by considering the LLL-projected bare interaction
\begin{equation}
    \gamma_0(\tilde \rr) = g_0(1 + \lambda_1 \tilde x^2 + \lambda_2\tilde  y^2) e^{-\tilde\rr^2 / 2}
    \label{eqn:anisotropic}
\end{equation}
with $g_0 <0$ (to generate attraction) and small $\lambda_{1,2}$ (to break rotation symmetry when $\lambda_1 \neq \lambda_2$).  Figure~\ref{fig:anisotropicPD} illustrates the numerically determined phase diagram obtained by integrating the flow equations for various initial $\lambda_{1,2}$ values.  Rotation symmetry is preserved only along the diagonal where $\lambda_1 = \lambda_2$.  The NFL clearly once again occupies a finite fraction of parameter space (blue region), and in particular enjoys resilience against appreciable explicit rotation symmetry breaking.   In the surrounding red region the NFL becomes unstable, though not towards superconductivity.  Instead the system flows to a \emph{repulsive} strong-coupling fixed point at which CDW states again emerge.  (Away from the diagonal the symmetry distinction between normal and nematic topological CDW's ceases to exist.) 

These findings suggest that reduced scattering from higher-moment conservation laws do \emph{not} fundamentally stabilize the NFL state.

\subsection{Spinful non-Fermi liquids}

Another natural modification to the symmetric LLL problem involves restoring spin as an active low-energy degree of freedom.  To this end we consider degenerate spin-up and spin-down LLL bands (ignoring Zeeman splitting). 
The interactions that we consider are captured by three coupling functions $(u,v,w)$ as
\begin{equation}
\begin{aligned}
    H_{\text{int}} &= \int_{z,\tau,\{k_y\}} \bigg{\{}\, u({k_y}_1 ,{k_y}_2)  \psi^\dagger_{\uparrow R,k_y + {k_y}_1  / 2} \psi^\dagger_{\uparrow L,k_y - {k_y}_1  / 2} \\
    & \qquad \quad  \times \psi_{\uparrow L,k_y - {k_y}_2 / 2} \psi_{\uparrow R,k_y + {k_y}_2 / 2} \\
    & \qquad + u \psi^\dagger_{\downarrow R} \psi^\dagger_{\downarrow L} \psi_{\downarrow L} \psi_{\downarrow R} \\
    & \qquad + \comm{v \psi^\dagger_{\uparrow R} \psi^\dagger_{\downarrow L} \psi_{\downarrow L} \psi_{\uparrow R} + (\uparrow  \leftrightarrow \downarrow ) } \\
    & \qquad + \comm{w \psi^\dagger_{\uparrow R} \psi^\dagger_{\downarrow L} \psi_{\uparrow L} \psi_{\downarrow R} + \text{h.c.} }\bigg{\}}. 
\end{aligned}
\label{eqn:interaction_spins}
\end{equation}
For brevity the last three lines are expressed schematically.  The $u$ terms represent the same interaction as in Eq.~\eqref{eqn:Sint} invoked separately for each spin; $v$ couples the densities for right- and left-movers with opposite spins; and $w$ backscatters the spin-up and spin-down electrons off each other.  
All interactions above preserve continuous spin-rotation symmetry about the $z$ direction.
We analyze this problem using functional RG and identify a tantalizing similarity between our numerical results and exact calculations for a one-dimensional Luttinger liquid obtained using bosonization.

\begin{figure}
    \centering
    \includegraphics[width=0.95\columnwidth]{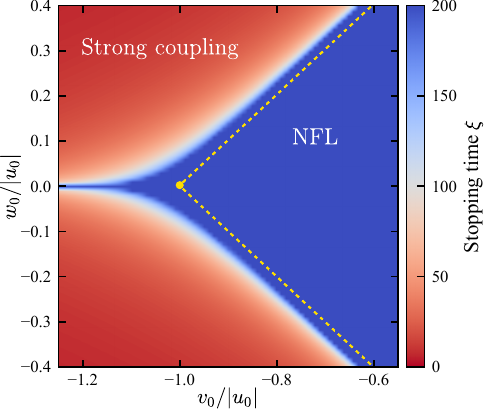}
    \caption{RG flow of the spinful interacting lowest Landau level. The phase boundary resembles the separatrix of the Berezinskii-Kosterlitz-Thouless flow equations (dotted line added for visualization) ending at $w_0 = 0, v_0 = u_0$ as predicted by a naive two-channel Luttinger liquid description (\ref{eqn:Hspin}). }
    \label{fig:PD2}
\end{figure}

Renormalization of the coupling functions follows from the one-loop flow equations
\begin{equation}
    \begin{aligned}
        \frac{d u}{d\xi} &= \frac{1}{2 \pi v_F}( u \cdot u -  u \ast u +  w^\ast \cdot w),  \\
        \frac{d v}{d\xi} &= \frac{1}{2 \pi v_F} (v \cdot v - v \ast v -  w^\ast \ast w),  \\
        \frac{d w}{d\xi} &= \frac{1}{2 \pi v_F} (- v \ast w - \, w \ast v + 2 u \cdot w).  \\
    \end{aligned}
    \label{eqn:flow_spins}
\end{equation}
We assume that the bare coupling functions are LLL-projected contact interactions, each with a single free parameter characterizing their initial strength as in \eqnref{eqn:g0g1} with the higher-derivative-term coefficient $g_1$ set to zero. Figure~\ref{fig:PD2} illustrates the phase diagram obtained from studying the flow numerically  assuming attractive bare $u<0$.  
In the fine-tuned $v = w = 0$ limit, the spins decouple, and although attractive $u$ interactions promote spin-triplet Cooper pairing at the mean-field level, a more accurate RG analysis reviewed earlier reveals that a NFL emerges instead \cite{Yakovenko1993}. Given that the system may be close to forming a superconductor, it seems plausible that adding the $w$ coupling further promotes Cooper pairing tendencies by enabling pairing between the two spin-species in the $s$-wave channel.  Indeed Fig.~\ref{fig:PD2} features a broad region in which the NFL phase is eventually destroyed by an instability triggered by $w$ flowing to strong coupling.  Whether the resulting phase is a superconductor, however, poses a subtle question that we return to shortly.

Remarkably, the phase diagram for our spinful LLL setup matches quantitatively with that of a strictly one-dimensional spinful fermionic system with analogous local interactions. The latter companion problem is described by a Hamiltonian 
\begin{widetext}
\begin{multline}
    H_{\text{1d}} = \int_{z}  \Bigg\{ \left[ \psi^\dagger_{\uparrow R}(-i v_F\d_z) \psi_{\uparrow R} - \psi^\dagger_{\uparrow L}(-i v_F \d_z) \psi_{\uparrow L} + \psi^\dagger_{\downarrow R}(-i v_F\d_z) \psi_{\downarrow R} - \psi^\dagger_{\downarrow L}(-i v_F \d_z) \psi_{\downarrow L}\right] \\ 
    + \left[u \psi^\dagger_{\uparrow R}\psi^\dagger_{\uparrow L}\psi_{\uparrow L}\psi_{\uparrow R} + \term{\uparrow \leftrightarrow \downarrow}\right] + \left[v \psi^\dagger_{\uparrow R}\psi^\dagger_{\downarrow L}\psi_{\downarrow L}\psi_{\uparrow R} + \term{\uparrow\leftrightarrow\downarrow}\right] + \left[w \psi^\dagger_{\uparrow R}\psi^\dagger_{\downarrow L}\psi_{\uparrow L}\psi_{\downarrow R} + \mathrm{h.c.}\right]  \Bigg\}.
\end{multline}
\end{widetext}
To emphasize connection to the original setup, we parametrized the one spatial dimension by $z$; note the identical structure of the $u,v,w$ terms compared to Eq.~\eqref{eqn:interaction_spins}.  

One can efficiently understand the impact of interactions in $H_{\rm 1d}$ using bosonization.  We use bosonized fields $\varphi_\sigma, \theta_\sigma$ related to the electron operators via $\psi_{\sigma,R/L} \sim e^{i(\varphi_\sigma \pm \theta_\sigma)}$ and follow conventions from Ref.~\onlinecite{FisherGlazman1997}.
The resulting bosonized Hamiltonian reads 
\begin{widetext}
\begin{multline}
    H_{\rm 1d} = \sum_{\sigma = \pm 1 \equiv \curly{\uparrow,\downarrow}}\int_{z}  \Bigg[ \frac{v_F}{2\pi}\term{(\d_z \varphi_\sigma)^2 + (\d_z\theta_\sigma)^2} - \frac{u}{(2\pi)^2} \d_z(\varphi_\sigma + \theta_\sigma)\d_z(\varphi_\sigma - \theta_\sigma) \\
    -\frac{v}{(2\pi)^2} \d_z(\varphi_\sigma + \theta_\sigma)\d_z(\varphi_{-\sigma} - \theta_{-\sigma}) + \frac{w}{a^2} \cos\term{2\theta_\uparrow - 2\theta_\downarrow}  \Bigg].
\end{multline}
\end{widetext}
where $a$ is a UV cutoff length scale. Introducing charge and spin variables $\varphi_c = (\varphi_\uparrow + \varphi_\downarrow)/\sqrt{2}, \varphi_s = (\varphi_\uparrow - \varphi_\downarrow)/\sqrt{2}$ (and similarly for $\theta$) yields the spin-charge separated Hamiltonian $H_{\rm 1d} = H_c + H_s$ with
\begin{align}
    H_c &= \int_z \frac{v_c}{2\pi} \comm{g_c (\d_z \varphi_c)^2 + g_c^{-1} (\d_z \theta_c)^2}, \label{eqn:Hcharge} \\
    H_s &= \int_z \bigg{\{}\frac{v_s}{2\pi} \comm{g_s (\d_z \varphi_s)^2 + g_s^{-1} (\d_z \theta_s)^2} 
    \nonumber \\
    &~~~~~~~~~+ \frac{2w} {a^2}\cos(2\sqrt{2}\theta_s)\bigg{\}}.
    \label{eqn:Hspin}
\end{align}
Here $v_c$ and $v_s$ are charge- and spin-sector velocities renormalized by $u,v$, while $g_c$ and $g_s$ are charge- and spin-sector Luttinger parameters.  Most importantly for our purposes, we have
\begin{equation}
    g_s = \sqrt{\frac{1 - \frac{u - v}{2\pi v_F}}{1 + \frac{u - v}{2\pi v_F}}}.
\end{equation}
From the bosonized description one deduces that the charge sector always remains gapless for weak interactions.  The spin-sector, however, shows a nontrivial phase transition generated by the $w$ term becoming relevant---in turn pinning $\theta_s$ to the minimum of the cosine and gapping the spin sector. 
In particular, the scaling dimension of $\cos(\alpha \theta_s)$ at the free-boson fixed point is given by  $\alpha^2 g_s / 4$, implying that the $w$ term becomes relevant for $g_s < 1$.  For attractive interactions with $u,v<0$, this condition translates to $|u|<|v|$.
The full flow equations display Berezenskii-Kosterlitz-Thouless (BKT) behavior.

Notably, this phase diagram matches the numerical RG of the three-dimensional LLL problem \emph{quantitatively} in the sense that the Luttinger liquid phase corresponds to the NFL region of parameter space (roughly given by $\abs{u} >\abs{v}$) where the functional RG does not flow to strong coupling. The LLL phase diagram also shows a striking resemblance to the BKT flow equations.
This remarkable correspondence seems unlikely to be a pathology of the techniques used to study the three-dimensional problem, which crudely resemble one dimension because of the flat in-plane dispersion.  Indeed we expect the functional RG analysis to be accurate in the limit of weak interactions with no other caveats.  
The connection discussed here hints that the NFL phase uncovered by Yakovenko may constitute a three-dimensional analog of a one-dimensional spinless Luttinger liquid, displaying phenomena such as spin-charge separation. 
(A Luttinger liquid description of the three-dimensional electron gas in the high-field limit has been invoked to explain various phenomena in Refs.~\onlinecite{BabichenkoKoslov1995,BiaginiMaslovGlazman2001, ZhangNagaosa2017},  although we expect that the NFL phase here is not adiabatically connected to an array of Luttinger liquids as visualized in these works.)  
 This connection further hints that the $w$-driven instability in Fig.~\ref{fig:PD2} may represent a spin-gapped but electronically gapless $U(1)_C$-preserving phase, rather than a superconductor.  The functional RG techniques we have employed do not readily distinguish these two possibilities.

Next we explore yet another perturbation---explicit breaking of translation symmetry---that manifestly does yield a superconducting state for the attractively interacting spinless LLL-projected system.

\section{Weyl Superconductor from explicitly broken translation symmetry} \label{sec:6breakingtranslation}

Let us recall the following two observations for the RG flow in the case of spin-polarized electrons in the LLL band:
\begin{enumerate}
    \item The flat Fermi surface leads to an extensive number of candidate CDW- and BCS-type nesting instabilities.  In particular, the flat dispersion yields a continuum of CDW nesting wavevectors $\qq_{\text{CDW}} = q \yhat + 2 k_F \zhat$, while Cooper pairs can potentially condense with total momentum $2\qq_{\rm BCS} = q \yhat$ ($q$ is arbitrary).  
    \item Although mean-field theory predicts a superconducting instability with attractive interactions, inclusion of the CDW diagram in the one-loop RG overrules this outcome---instead driving a flow to a NFL state.
\end{enumerate}
Since competition from the CDW ostensibly prevents Cooper pairs from condensing, a natural route to superconductivity is to obliterate the Fermi surface CDW nesting condition (while retaining a tendency towards pairing).  
As noted by Yakovenko \cite{Yakovenko1993}, such a scenario arises upon explicitly breaking translation symmetry in the plane perpendicular to the magnetic field. We now pursue this route by adding a weak periodic potential 
\begin{equation}
    V(x) = -V_0 \cos(2 \pi x / \lambda)
    \label{eq:V}
\end{equation} 
along one of the directions in the transverse plane.  Importantly, the potential preserves the effective dipole conservation symmetry along $x$ but \emph{not} along $y$; recall the discussion from the start of Sec.~\ref{sec:attractiveinstabilities}.  We temporarily assume $\lambda \gg l_B$ for simplicity. In this regime we can apply degenerate perturbation theory after LLL projection to quantify how the periodic potential modifies the single-particle dispersion. Since  translation symmetry persists along the $y$ direction, $k_y$ remains a good quantum number and the single-particle eigenstates remain the same. The eigenenergies are modified, however, and become $k_y$-dependent; to leading order in $l_B/\lambda$ one obtains
\begin{equation}
    \epsilon_{k_y,k_z} = \frac{k_z^2}{2m}  - V_0 
    \cos\term{\frac{2 \pi k_y l_B^2}{\lambda}} - \mu.
\end{equation}
The $V_0$ correction simply arises from replacing $x$ in Eq.~\eqref{eq:V} with the mean position $k_y l_B^2$ in a given LLL state---which is sensible because the periodic potential varies weakly over the spread of the wavefunction in the regime under consideration. 

\begin{figure}
    \centering
    \includegraphics[width=1.0\linewidth]{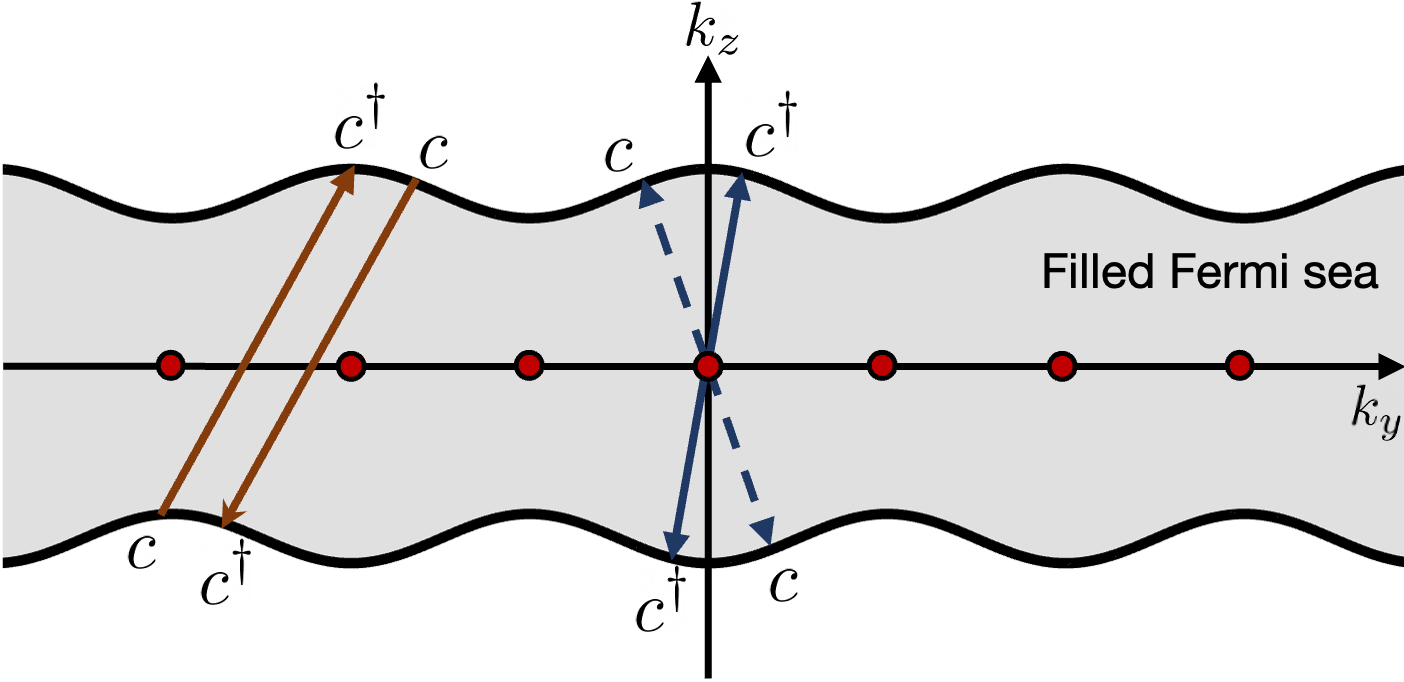}
    \caption{Curved Fermi surface resulting from the periodic potential $V(x)$ from Eq.~\eqref{eq:V}. Compared to the flat Fermi surface in Fig.~\ref{fig:Flat_FS}, where a continuum of CDW and BCS nesting conditions arise, only a discrete set of potential instabilities remain when curvature is present.  Orange arrows indicate the shortest remaining CDW nesting wavevectors, while red circles indicate the remaining discrete $\qq_{\rm BCS}$ values that determine the dominant Cooper-pair momenta.  Locality renders CDW instabilities ineffective, allowing superconductivity to emerge from weak attractive interactions.     }
    \label{fig:Curved_FS}
\end{figure}

Figure~\ref{fig:Curved_FS} illustrates the resulting Fermi surface---which is now curved in such a way that most (but not all) CDW and BCS nesting instabilities are destroyed.  Residual candidate CDW nesting instabilities now connect states with $k_y$ momenta differing by large \emph{discrete} values $\Delta k_n = (\lambda/l_B^2)(n+1/2)$ 
for $n \in \mathbb{Z}$; see orange arrows in Fig.~\ref{fig:Curved_FS} for an example.  These CDW processes can be safely ignored due to locality: any bare short-range interaction that would promote this channel is exponentially suppressed in $(\lambda/\l_B)^2$, as the participating $k_y$ states exponentially localize at distant $x$ positions separated by $\Delta x_n = \Delta k_n l_B^2$.  The residual candidate BCS instabilities similarly involve Cooper pairs carrying discrete total momentum $2\qq_{\rm BCS} = n (\lambda/l_B^{2})  \yhat, n \in \mathbb{Z}$ (red circles in Fig.~\ref{fig:Curved_FS} illustrate the set of $\qq_{\rm BCS}$ vectors).  Crucially, states with nearby $k_y$'s contribute to candidate BCS instabilities with some given discrete Cooper-pair momentum---hence no suppression from locality appears here contrary to the CDW case.  To an excellent approximation we can therefore ignore competition from the leftover CDW instabilities and focus exclusively on the BCS sector.  

\begin{figure}
    \centering
    \includegraphics[width=.95\linewidth]{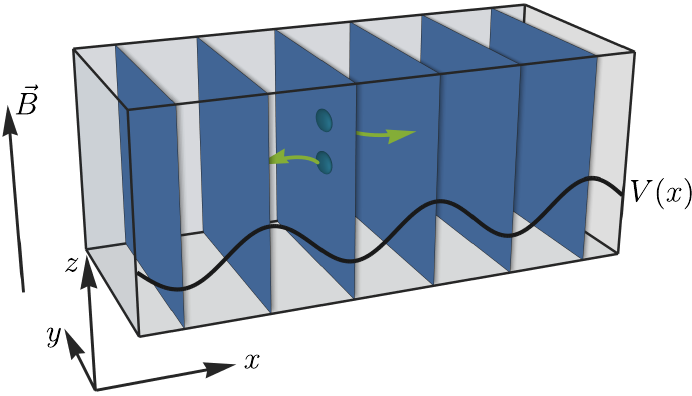}
    \caption{Schematic of layered bulk superconductor arising from islands that localize to the minima and maxima of the periodic potential $V(x)$.  Attractive interactions within each island catalyze superconductivity with phase coherence along both the $y$ and $z$ directions. The blue spheres denote Cooper pairs, and the lowest order coupling between the layers is given by a coordinated Cooper pair hopping (see \eqnref{eqn:BEI}) that conserves the dipole moment along $x$.  Such processes do \emph{not} generate phase stiffness transverse to the islands.  The bulk layered superconductor thus exhibits highly anisotropic transport---supporting supercurrents in the $y$ and $z$ directions yet insulating along the $x$ direction. }
    \label{fig:Layered_SC}
\end{figure}

Locality does still play a role even in the simplified problem that neglects CDW processes.  Namely, $k_y$ states contributing appreciably to Cooper pairs carrying different discrete total momentum localize to different positions in real space.  We can, as an initial foray, therefore view the periodic potential as breaking the system up into independent `islands', oriented along the $(y,z)$ plane and localized to the maxima and minima of $V(x)$ at positions $x_j = \lambda j/2$ with $j \in \mathbb{Z}$ the island index; see Fig.~\ref{fig:Layered_SC}. 
Using the usual relation $x \sim k_y l_B^2$, electrons in layer $j$ carry $y$-direction momentum given approximately by $k_{y,j} = (\lambda/l_B^2)j/2$. 
Unencumbered by competition from CDW processes, within each island attractive interactions generate superconductivity with island-dependent Cooper-pair momentum $(\lambda/l_B^2)j\hat{\boldsymbol{y}}$ and non-zero phase stiffness along both the $y$ and $z$ directions.  To uncover physical properties of the superconductor, below we pursue a mean-field treatment of pairing for a single island, then restore residual (Josephson) coupling between islands, and finally extract the fermionic quasiparticle spectrum from the full three-dimensional problem in the limit where the islands couple strongly.

\subsection{BCS mean-field theory for a superconducting island} 

\begin{figure}
    \centering
    \includegraphics[width=0.4\columnwidth]{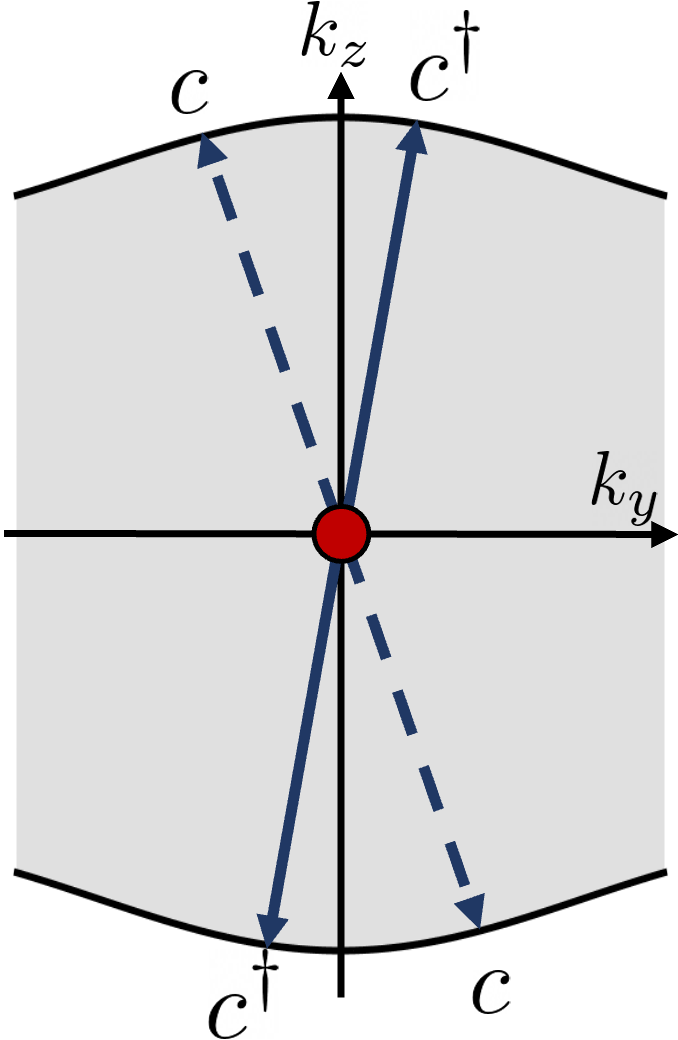}
    \caption{Patch of the curved Fermi surface from Fig.~\ref{fig:Curved_FS} relevant for superconductivity in a single island.  Arrows indicate the  associated Cooper instability resulting from attraction. States far from the island in the $k_y$ direction do not participate appreciably in pairing due to locality.}
    \label{fig:FS_single_island}
\end{figure}

Let us begin by zooming in on an individual superconducting island.  Figure~\ref{fig:FS_single_island} depicts the operative slice of the Fermi surface. 
Without loss of generality, we focus on island $j = 0$ located at position $x = 0$, so that relevant momenta are centered around $k_y$ = 0.  All other islands are related via magnetic translation symmetry. Appendix~\ref{app:BCSMF} carries out the BCS mean-field calculation, yielding three main takeaways:

\begin{enumerate}
    \item The Boguliubov quasiparticle gap $\Delta_{k_y}$ on the Fermi surface decays exponentially as a function of $k_y$: 
    \begin{equation}
        \Delta_{k_y} = \Delta_0 e^{-2 k_y^2 l_B^2},
        \label{eq:Delta_ky}
    \end{equation}
    where $\Delta_0$ is the characteristic pairing field.  The decay originates from the fact that  resonant Cooper pairs on the Fermi surface with momenta $k_y$ and $-k_y$ become increasingly spatially separated as $k_y$ increases; correspondingly, the short-range attractive interactions that pair them become suppressed by locality.
    
    \item The critical temperature is given by
    \begin{equation}
        T_c \approx 1.36 \hbar \omega_D \exp\term{\frac{-2 \sqrt{2}}{g(2 \pi v_F \sqrt{2 \pi l_B^2})^{-1}}}.
    \end{equation}
    Here, $g$ is the interaction strength and $\omega_D$ is the Debye frequency, a UV cutoff associated to the interaction. This result resembles that of BCS theory where $\sim (2 \pi v_F \sqrt{2 \pi l_B^2})^{-1}$ plays the role of an effective density of states at the Fermi level. 

    \item Denoting the `gap' by the energy scale $\Delta_0$, we obtain the universal gap-to-$T_c$ ratio 
    \begin{equation}
        \frac{\Delta_0}{k_B T_c} = 2.26,
    \end{equation}
    which as expected differs quantitatively from the standard BCS result. 
\end{enumerate}

\subsection{Josephson coupling between superconducting islands}
\label{sec:Bulk_Superconductor}
To lowest order, locality justifies neglecting the interactions between islands, though we now wish to go beyond that approximation.  What are the leading symmetry-allowed inter-island Josephson couplings?  The system with the external periodic potential preserves translation symmetry along $y$, and so all interactions must conserve total $k_y$ momentum.  Recall that, crucially, Cooper pairs in island $j$ carry an island-dependent $y$-direction momentum $(\lambda/l_B^2)j$.  Momentum conservation thereby forbids conventional Josephson coupling mediated by Cooper pairs hopping between any pair of islands.
The lowest order symmetry-allowed Josephson-type coupling instead arises from a coordinated dipole-conserving hopping process where two Cooper pairs from island $j$ interact, sending one to $j+1$ and the other to $j-1$ (see \figref{fig:Layered_SC}). 
In terms of a coarse-grained superconductor phase variable\footnote{Note that $\varphi_j$ is not the microscopic phase variable associated to the superconductor, but a coarse grained one where the short-wavelength oscillations of the phase are taken out. This is similar to the Tkachenko field of the LLL superfluid described in Ref.~\onlinecite{Du2024}.} $\varphi_j$ for island $j$, such processes lead to the nontrivial Josephson Hamiltonian
\begin{equation} \label{eqn:BEI}
    H_{\text{J}} = J\sum_j\cos(2 \varphi_j - \varphi_{j+1} - \varphi_{j-1}).
\end{equation}
The form above implies that the layered superconducting state exhibits no phase stiffness along the $x$ direction, transverse to the islands.  Indeed, a linearly increasing $\delta\varphi_j \sim j$ phase profile costs no energy. Such interactions have been discussed in dipole-conserving bosonic systems~\cite{LakeHermeleSenthil2022}, and it is known that the ordered state is a Bose-Einstein insulator (BEI) that spontaneously breaks $U(1)$ charge conservation symmetry yet insulates due to dipole conservation symmetry that precludes charge transport. In our problem, dipole conservation along the $x$ direction originates from magnetic translation symmetry in the $y$ direction. An interesting consequence of this microscopic origin is that dipole conservation is broken near samples boundaries, enabling superconducting edge currents in the $x$ direction. 
We leave a detailed analysis of such boundary transport physics to future work.

\subsection{Full theory of the bulk mean-field superconductor}
\label{sec:BulkSuperconductor}

Incorporating correlated transfer of bosons between islands sufficed for inferring the absence of bulk supercurrents transverse to the islands, but does not reveal the generic nature of the fermionic quasiparticle spectrum.  The quasiparticle spectrum is particularly subtle given exponential decay of the intra-island quasiparticle gap with momentum $k_y$; recall Eq.~\eqref{eq:Delta_ky}.  Correspondingly, even nominally small inter-island fermionic couplings could play a decisive role in governing the gap structure for the full three-dimensional problem.  

To address this issue, we now revisit the BCS mean-field analysis without invoking the $\lambda \gg l_B$ condition imposed previously (though we still assume that higher Landau levels remain inert).  Attacking the problem from weakly interacting superconducting islands then becomes inappropriate; instead, a treatment of the full bulk theory is required.  
The mean-field Hamiltonian consists of two parts,
$H = H_{\text{KE}} + H_{\text{Pairing}},$
respectively encoding kinetic energy along the field direction and Cooper pairing:
\begin{widetext}
    \begin{align}
        H_{\text{KE}} &= \int_{k_y,q_z} v_F q_z\term{\psi^\dagger_{R, k_y,q_z}\psi_{R,k_y,q_z} - \psi^\dagger_{L,k_y,q_z}\psi_{L,k_y,q_z}}, 
        \label{eq:HKE} \\
        H_\text{Pair} &= \sum_{K_y \in \frac{\lambda}{2 l_B^2} \mathbb{Z}} \int_{k_y,q_z} \left[\Delta(K_y, k_y) \psi^\dagger_{R,K_y + k_y,q_z} \psi^\dagger_{L,K_y-k_y,-q_z} + \text{h.c.}\right].
        \label{eq:Hpair}
    \end{align}
\end{widetext}
The pairing term arises from a mean-field decoupling of the attractive interaction that flows to strong coupling, assisted by the applied periodic potential.  In principle, we should solve for the pairing field $\Delta(K_y,k_y)$ self-consistently, though here we pursue a simplified symmetry-based approach that we expect captures universal features.  

The $K_y \in (\lambda/l_B^2)/2 \times \mathbb{Z}$ parameter summed over in $H_\text{Pair}$ indicates the center of mass momentum of the Cooper pair created by the pair field.  (Hereafter we use uppercase letters exclusively for discrete center-of-mass momenta.) One can roughly associate $K_y$'s with the condensates residing within each superconducting island.
Because $k_y$ is integrated over all momenta, however, Cooper pairs can arise from electrons residing in different islands even for a given fixed $K_y$ value.  Locality nevertheless continues to dictate that $\Delta(K_y, k_y)$ decays exponentially in $k_y$. Imposing magnetic translation symmetry allows the decomposition $\Delta(K_y,k_y) = e^{i \Phi(K_y)} \Delta_0(k_y)$.  Reflection symmetry $\mathcal{M}_z$ further enforces that $\Delta_0$ is either even [$\Delta_0(k_y) = \Delta_0(-k_y)$] or odd [$\Delta_0(k_y) = -\Delta_0(-k_y)$].  We focus on the even branch since the dominant pairing arises for momenta near $k_y = 0$.  
For analytic tractability we will assume $\Phi(K_y) = 0$, commenting on other choices below. 

\begin{figure}
    \centering
    \includegraphics[width=1.0\linewidth]{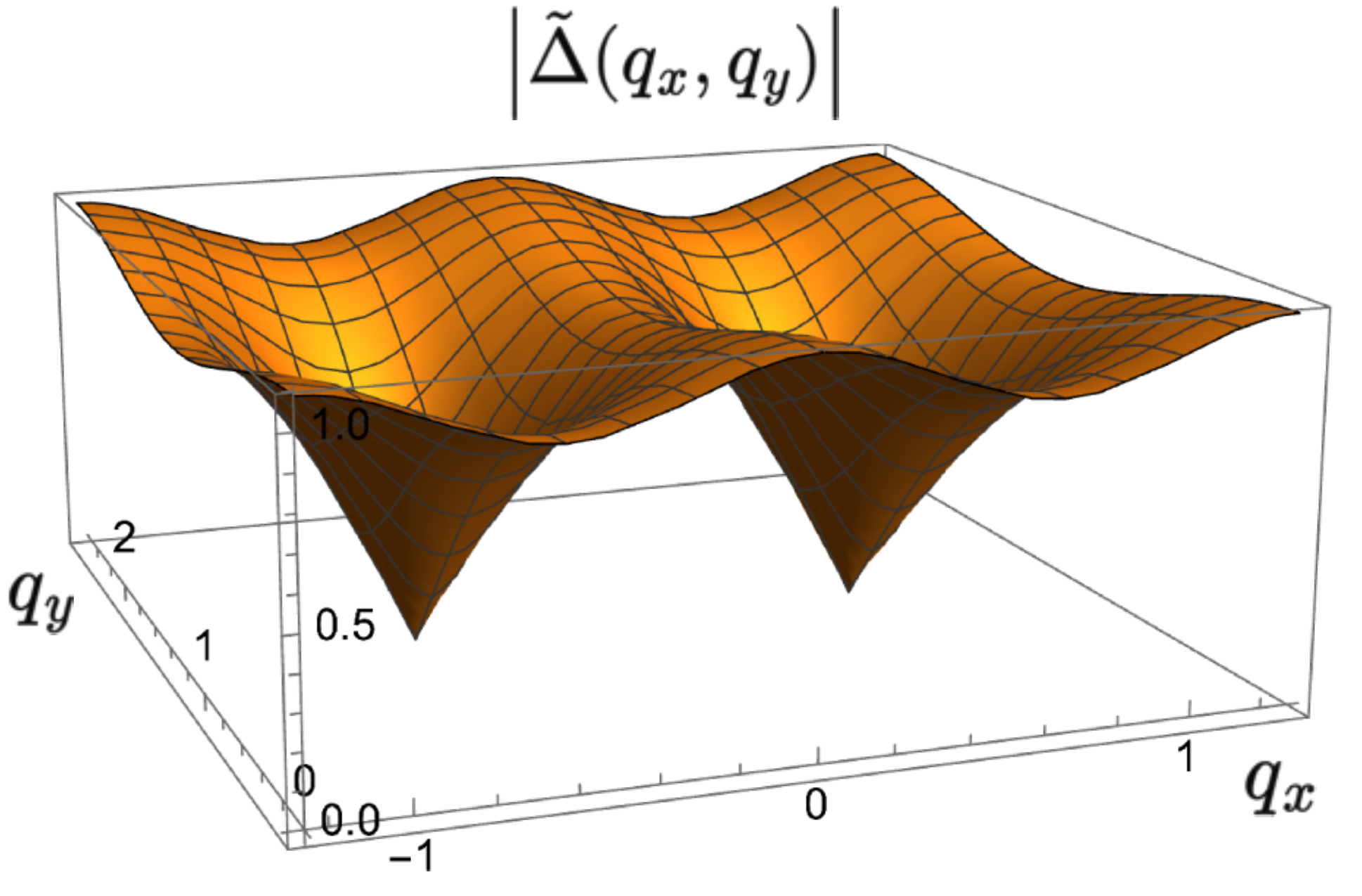}
    \caption{Absolute value of the pairing function $\tilde\Delta(q_x,q_y)$ plotted for $\Delta_0(k_y) = \exp(-k_y^2 l_B^2 /2), l_B=1$, and $\lambda = 5.$ The two zeros correspond to the pair of Weyl nodes arising in the spectrum [Eq.~\eqref{eqn:Bspec}] at $q_z = 0$.  } 
    \label{fig:DeltaTilde}
\end{figure}

Discrete translation symmetry along the $x$ direction implies that the mean-field Hamiltonian can be recast as a lattice problem for `sites' $K_y$ with lattice constant $\lambda/(2l_B^2)$.  To this end we first equivalently express Eqs.~\eqref{eq:HKE} and \eqref{eq:Hpair} as
\begin{widetext}
\begin{align}
\begin{split}
    H_{\rm KE} &= \sum_{K_y \in \frac{\lambda}{2 l_B^2} \mathbb{Z}}\int_0^{\frac{\lambda}{2 l_B^2}}dq_y \int dq_z v_F q_z\term{\psi^\dagger_{R, K_y+q_y,q_z}\psi_{R, K_y+q_y,q_z} - \psi_{L,K_y-q_y,-q_z} \psi^\dagger_{L,K_y-q_y,-q_z} } \\ 
    H_{\rm Pair} &= \sum_{K_y,K_y'\in \frac{\lambda}{2 l_B^2} \mathbb{Z}} \int_0^{\frac{\lambda}{2 l_B^2}}dq_y \int dq_z\left[\Delta_0(K'_y+q_y) \psi^\dagger_{R,K_y+K'_y +q_y,q_z} \psi^\dagger_{L,K_y-K'_y -q_y,-q_z} + \text{h.c.}\right] .
\end{split}
\label{eqn:single_particle_H}
\end{align}
\end{widetext}
Next we Fourier transform, trading in the `position' $K_y$ for a conjugate `momentum' that we denote by $q_x$:
\begin{align}
    \psi_{R,K_y+q_y,q_z} = \frac{1}{\sqrt{N}} \sum_{q_x} e^{i q_x K_y l_B^2} \psi_{R,{\qq}}
\end{align}
and similarly for left movers;  here $N$ is the number of $K_y$ points, $\qq = (q_x,q_y,q_z)$, and $q_x \sim q_x + 2 \pi (\lambda / 2)^{-1}$ is summed over the Brillouin zone.  Finally, in terms of a two-component spinor $\phi^T_{\qq} = [\psi_{R,\qq}~ \psi^\dagger_{L,-\qq}]$ and the lattice Fourier transform of the pairing potential 
$\tilde \Delta_0(q_x,q_y) = \sum_{K_y} e^{-i K_y 2 q_x l_B^2} \Delta_0(K_y + q_y)$, the Hamiltonian takes a standard Bogoliubov-de Gennes form 
\begin{align}
    H = \sum_{q_x} \int_0^{\frac{\lambda}{2 l_B^2}}\frac{dq_y}{2\pi} \int \frac{dq_z}{2\pi} \phi^\dagger_{\qq} \mathcal{H}_{\qq} \phi_{\qq}
\end{align}
with
\begin{align}
    \mathcal{H}_{\qq} = \begin{pmatrix}
        v_F q_z & \tilde \Delta_0(q_x,q_y) \\
        \tilde \Delta_0^\ast(q_x,q_y) & -v_F q_z
    \end{pmatrix}.
\end{align}
Fermionic quasiparticle excitation energies immediately follow as
\begin{equation}\label{eqn:Bspec}
    \xi_{\qq} = \sqrt{v_F^2 q_z^2 + |\tilde \Delta(q_x,q_y)|^2}.
\end{equation}
The spectrum becomes gapless when $q_z = 0$ and $\tilde \Delta(q_x,q_y) = 0$, the latter vanishing when $q_x = \pm (\pi /2) (\lambda / 2)^{-1}$ and $q_y = (\lambda/4)l_B^{-2}$ as $\Delta_0(k_y)$ is an even function; see \figref{fig:DeltaTilde} for a typical $|{\tilde\Delta}(q_x,q_y)|$ profile.  The superconductor therefore hosts two complex Weyl fermions at low energy. 

In \appref{app:CooperProfile}, we show that this mean-field state corresponds to a superconductor with a condensate order parameter that has an array of vortices forming a rectangular lattice (an Abrikosov vortex lattice) in the $xy$ plane, with lattice constant $\lambda/2$ in the $x$ direction  and $\pi l_B^2 / \lambda$ in the $y$ direction. We note that the aspect ratio of the unit cell is set by the periodicity of the external potential we put in by hand. The unit cell area of the vortex lattice, however, is fixed to be $\pi l_B^2$ by the magnetic field (as we have a charge 2 superconductor), so there are two vortices per magnetic flux quantum (defined with respect to the electronic charge). One could obtain other configurations of the vortex lattice by changing the phase of the pairing function $\Delta(K_y, k_y)$ in a $K_y$ dependent manner.

One can intuitively understand the bosonic sector of this theory using an effective Ginzburg-Landau description if we assume the island-dependent pairing function factorizes as $\Delta(K_y,k_y) = \Delta_0 (k_y) \Delta_{\text{island}}(K_y)$ (earlier we assumed it is independent of $K_y$). Then, one can identify this mean-field state with a condensate Ginzburg-Landau wavefunction $\Psi_{\text{GL}}(\rr)$ in the charge $2e$ LLL 
\begin{equation}
    \Psi_{\text{GL}}(\rr) \propto \sum_{K_y \in \lambda / (2 l_B^2) \mathbb{Z}} \Delta_{\text{island}}(K_y)  e^{2i K_y y - (x-K_yl_B^2)^2/l_B^2}.
\end{equation}
Let us specialize to the case $\Delta_{\text{island}}(K_y) = 1$ and look at the contribution to $\Psi_{\text{GL}}(\rr)$ from two neighboring $K_y$'s. They have the same Gaussian spread offset in the $x$ direction by $\lambda / 2$, but their phases wind at different rates in the $y$ direction. Specifically, there is a relative winding of $e^{i \lambda y /l_B^2}$. The resulting constructive and destructive interference in $\Psi_{\text{GL}}(\rr)$ produces zeros (vortices) between the two islands spaced at equal intervals of $2\pi l_B^2/\lambda$ in the $y$ direction. Since at $y = 0$ all the islands are in phase, this implies that the resulting vortex lattice will be rectangular with the other lattice vector being $\lambda /2 \hat{\boldsymbol{x}}$. 

 Using the above intuition, we also see that for the vortices to be offset in the $y$ direction between each neighboring pair of layers (which is presumably does to lower energy~\cite{ABRIKOSOV1957}), we want $\Delta_{\text{island}}(K_y)$ to follow the pattern $+1,+1,-1,-1,\dots$ for neighboring islands as it is the \emph{relative phase} between the islands that dictates the location of the vortices in the $y$ direction. 
 The case of the triangular lattice and associated Boguliubov spectrum is discussed in \appref{app:tr_vortex_lattice}. The true ground state configuration of the vortex lattice is determined by energetics, which is beyond the scope of this work.

\section{Discussion}
We have revisited the phase diagram of the three-dimensional weakly interacting electron gas in a strong magnetic field with various physically motivated deformations. Yakovenko~\cite{Yakovenko1993} first attacked this problem using a functional RG procedure and studied the phase diagram for LLL-projected attractive and repulsive contact interactions. From general principles, we argued that the contact interaction is not singled out under RG, higher in-plane derivative interactions are equally relevant, and the space of marginal perturbations of the free theory is infinite-dimensional.  We studied a class of these perturbations numerically and found that there are three distinct phases that appear generically in the limit of weak local interactions---the normal topological CDW, its nematic counterpart, and a NFL.  We studied the first two phases within a Hartree-Fock treatment of the renormalized interaction and suggested experiments to detect them. 

An intriguing puzzle arising from the above study was the apparent absence of any attractive (presumably superconducting) instabilities. We reviewed the relationship between spatial symmetries and higher moment conservation in the LLL, and the resulting modified Hohenberg-Mermin-Wagner theorems that nevertheless do not explain this observation. To gain more insight, we modified the problem by 
\begin{enumerate}
    \item Changing the Landau level index: we found a qualitatively similar phase diagram featuring a resilient NFL phase and lacking any attractive instabilities.  Evidently the special structure of the LLL wavefunctions is not essential for the NFL.
    \item Breaking rotational symmetry: similarly, the phase diagram exhibited  a robust NFL phase but no attractive instabilities.  Here we learned that the NFL also does not originate from the reduced phase space for scattering that stems from rotational symmetry preserving the trace quadrupole moment of charge.  
    \item Adding a spin index and allowing the two species to interact: once again the NFL appears stable, but over a regime of interaction parameters  becomes unstable when a term that backscatters spin-up and spin-down electrons off of each other flows to strong coupling.  We observed that the phase diagram appears to match that of a spinful one-dimensional electron liquid.  That is, the Luttinger liquid phase (with gapless charge and spin sectors) and the NFL appear under quantitatively similar interaction parameter regimes.  This correspondence suggests that the NFL may exhibit Luttinger-liquid phenomena such as spin-charge separation; it further hints that when the NFL phase becomes unstable due to the backscattering term becoming relevant, the system enters a three-dimensional analog of a Luther-Emery liquid (with a gapless charge sector but gapped spin sector).  Putting these speculations on firmer footing poses a very interesting open problem.  
\end{enumerate}

All of the above modifications preserved the pair of effective dipole conservation symmetries exhibited within a projected Landau level.  We also examined a deformation that explicitly violates one of the two dipole conservations: addition of a periodic potential along the $x$ direction, transverse to the magnetic field.  The periodic potential effectively killed the density wave channel; in turn, weak attractive interactions (compared to both the periodic potential and Fermi energy) no longer generated a NFL but rather triggered superconductivity.  Our results therefore collectively suggest that \emph{stability of the NFL phase is tied fundamentally to preservation of dipole conservation symmetries}.  One caveat is that we did not analyze the problem with attraction that is strong compared to the periodic potential.  We thus can not rule out the (somewhat unnatural) possibility that superconductivity emerging at weak attraction eventually gives way to a NFL at strong attraction.  

The superconductor itself also hosts remarkable properties in part descending from the lone dipole conservation symmetry preserved by the periodic potential. 
We argued that the low-temperature phase realizes a layered superconductor that supports supercurrents parallel to the layers, but behaves as a Bose-Einstein insulator in between the layers.  Indeed the remaining dipole symmetry prevents transport of charge between layers and heavily restricts the allowed interlayer Josephson couplings.  Interestingly, near the sample boundary, this residual symmetry is broken, and interlayer charge transport is allowed. Such superconducting edge transport is an example of the rich phenomenology of this phase that warrants further investigation.  The superconductor also hosts a nontrivial Bogoliubov quasiparticle spectrum featuring a pair of Weyl nodes.  
These Weyl nodes are topologically protected---similar to the Boguliubov Weyl nodes that appear when Fermi surfaces with Chern number pair with each other~\cite{LiHaldane2018}---and imply that the surface will support Bogoliubov-Fermi arcs.  Possible relation between the surface Fermi arcs and edge transport phenomenology highlighted above raises yet another interesting open question.  

\begin{acknowledgments}

We thank Ganapathy Baskaran, Anton Kapustin, and Leo Radzihovsky for helpful discussions.  This work was supported by the Gordon and Betty Moore Foundation’s EPiQS Initiative, Grant GBMF8682, and Caltech's Institute for Quantum Information and Matter, an NSF Physics Frontiers Center (Grant No. PHY-2317110).

\end{acknowledgments}

\bibliography{fracton}

\clearpage
\onecolumngrid

\appendix

\section{Symmetry allowed LLL projected interactions}
\label{App:SymmetryAllowedInteractions}
In this section, we characterize the general form of all symmetry-allowed interaction terms in the LLL-projected Hamiltonian for spinless right- and left-moving fermions. We can work with the right and left-moving fields in the LLL as they are resolved by $\partial_z$ acting on the bare electron field in the low-energy limit. It is useful to write the coupling function in the interaction term (\ref{eqn:Sint}) in the most general form
\begin{equation} \label{eqn:GeneralInteraction}
    S_{\text{int}} = \int_{\{k_y, q_z\}} d\tau \;h({k_y}_1,{k_y}_2, {k_y}_3, {k_y}_4)   \psi^\dagger_{R,{k_y}_1 } \psi^\dagger_{L,{k_y}_2} 
    \psi_{L, {k_y}_3} \psi_{R,{k_y}_4} 
     \delta(\Sigma q_z).
\end{equation}

We begin by identifying the symmetries in our problem and the explain the consequences for the interaction ($\alpha = R/L \equiv \pm 1$). 
\begin{itemize}
    \item Spacetime symmetries:
    \begin{enumerate}
        \item Continuous (magnetic) translation symmetry:
        \begin{enumerate}
            \item Along $x$.
        \begin{equation}
            \psi_{\alpha, k_y, q_z} \mapsto \psi_{\alpha, k_y + a_x l_B^{-2}, q_z}
        \end{equation}
        This symmetry implies that $h(\{ k_y \})$ in \eqnref{eqn:GeneralInteraction} can only depend on momentum differences, i.~e. $h(\{ k_y \}) = h(\{ k_y - k_0 \})$.
        \item Along $y$.
        \begin{equation}
            \psi_{\alpha, k_y, q_z} \mapsto e^{i k_y a_y} \psi_{\alpha, k_y, q_z}
        \end{equation}
        This symmetry implies momentum conservation $ k_{y,1}+k_{y,2} - k_{y,3} - k_{y,4} = 0$ in \eqnref{eqn:GeneralInteraction}.
        \end{enumerate}
        Note that the two translation symmetries do not commute, but satisfy the magnetic translation algebra as the field $\psi$ is charged.
        \item $SO(2) \equiv U(1)$ rotation symmetry in the $xy$ plane.
        \begin{equation}
            \psi_\alpha(\boldsymbol{r}) \mapsto U(\theta)\, \psi_{k_y, q_z}\, U(\theta)^{-1}
            = \int \! \frac{dk_y'}{2\pi}\; K_\theta (k_y,k_y')\, \psi_{k_y', q_z},
        \end{equation}
        where 
        \begin{equation}
            K_\theta(k,k') = \sqrt{\frac{l_B^2}{2\pi i \, {\sin \theta}}}\;
            \exp\!\left\{ \frac{i l_B^2}{2 \sin \theta}\left[ (k^2+k'^2)\cos \theta - 2 k k' \right] \right\}.
        \end{equation}
        To derive this, we used the generator of rotations $\hat{L} = \hx^2 + \hy^2$ in the LLL, whose eigenstates are the circular Landau level orbitals, familiar from solving the Landau level problem in symmetric gauge. The action of the rotation operator in the $k_y$ basis is given by 
        \begin{align}
            \bra{k_y} U(\theta) \ket{k_y'} &= \sum_{m,m' = 0}^\infty \braket{k_y}{m}  \bra{m} U(\theta) \ket{m'} \braket{m'}{k_y'}\\
            &= \sum_{m = 0}^\infty e^{-i m \theta} \braket{k_y}{m} \braket{m}{k_y'} \\
            &= K_\theta(k_y,k_y').
        \end{align}
        where the last step is the Mehler kernel, the real-space propagator for the harmonic oscillator. 

        This symmetry implies that $h(\{ k_y \})$ in \eqnref{eqn:GeneralInteraction} must satisfy
        \begin{align}
            &\int \prod_{i=1}^4 \frac{dk_{y,i}'}{2\pi} K_\theta(k_{y,1},k_{y,1}') K_\theta(k_{y,2},k_{y,2}') K_\theta^\ast(k_{y,3},k_{y,3}') K_\theta^\ast(k_{y,4},k_{y,4}') \nonumber \\
            &\qquad \times h(k_{y,1}',k_{y,2}',k_{y,3}',k_{y,4}') = h(k_{y,1},k_{y,2},k_{y,3},k_{y,4}) .
        \end{align}
        Using the magnetic translation symmetries, we know that $h(\{ k_y \}) = h(k_{y,1}-k_{y,2}, k_{y,4}-k_{y,3})$ and $k_{y,1} + k_{y,2} = k_{y,3}+ k_{y,4}$. Now, it is useful to recall that $K_\theta$ is the propagator of the simple harmonic oscillator with Hamiltonian $H = \frac{\theta}{2} \term{-\partial_x^2 + x^2}$.  Therefore, we can write the constraint from rotation symmetry as
        \begin{equation}
            e^{i\frac{\theta}{2} \term{-\partial_{x_1}^2 + x_1^2  -\partial_{x_2}^2 + x_2^2 + \partial_{x_3}^2 - x_3^2  +\partial_{x_4}^2 - x_4^2}} h(x_1-x_2,x_4-x_3) = h(x_1 - x_2,x_4 - x_3) .
        \end{equation}
        Defining $x_1^\pm \coloneqq (x_1 \pm x_2)/\sqrt{2},x_2^\pm \coloneqq (x_4 \pm x_3)/\sqrt{2}$. In this notation, we have 
        \begin{equation}
            e^{i\frac{\theta}{2} \term{-\partial_{x^+_1}^2 + (x^+_1)^2  -\partial_{x_1^-}^2 + (x_1^-)^2 + \partial_{x_2^+}^2 - (x_2^+)^2  +\partial_{x_2^-}^2 - (x_2^-)^2}} h(\sqrt{2} x_1^-,\sqrt{2} x_2^-) = h(\sqrt{2} x_1^-,\sqrt{2} x_2^-)
        \end{equation}
        due to the linearity of the simple harmonic oscillator. Here we note that the derivatives in $x_{1,2}^+$ become zero and $(x_1^+)^2 - (x_2^+)^2 = 0$ by momentum conservation. Therefore, the above equation simplifies to
        \begin{equation}
            e^{i\frac{\theta}{2} \term{ -\partial_{x_1^-}^2 + (x_1^-)^2  +\partial_{x_2^-}^2 - (x_2^-)^2}} h(\sqrt{2} x_1^-,\sqrt{2} x_2^-) = h(\sqrt{2} x_1^-,\sqrt{2} x_2^-).
        \end{equation}
        Defining $x = \frac{x_1^- + x_2^-}{\sqrt{2}}, x' = \frac{x_2^- - x_1^-}{\sqrt{2}}$, we see that this is equivalent to 
        \begin{equation}
            e^{i\theta \term{-\partial_{x} \partial_{x'} + x x'}} h\term{{x+x'},{x'-x}} = h\term{{x+x'},x'-x} .
        \end{equation}
        Recalling the definition of the $\lambda$ coupling function in \eqnref{eqn:gamma_rep}, this is equivalent to the constraint
        \begin{equation}
            e^{i\theta \term{-\partial_{x} \partial_{x'} + x x'}} \lambda(x,x') = \lambda(x,x') .
        \end{equation}
        Applying a Fourier transform to take $x' \to y$, we see that the constraint becomes
        \begin{equation}
            e^{i\theta \term{ -i x \partial_y + i y \partial_x}} {\gamma}(x,y) = {\gamma}(x,y)
        \end{equation}
        where ${\gamma}(x,y)$ is the Fourier transform of $\lambda(x,x')$ in the second variable (defined in \eqnref{eqn:gamma_rep}). This is simply the generator of rotations in the $xy$ plane, so we see that rotation symmetry implies that ${\gamma}(x,y)$ is rotationally symmetric.

        \item Translation symmetry along $z$. The field transforms as 
        \begin{equation}
            \psi_{\alpha,k_y,q_z} \mapsto e^{i q_z a_z}\psi_{\alpha,k_y,q_z}
        \end{equation}
        This enforces the conservation rule $q_{z,1} + q_{z,2} = q_{z,3} + q_{z,4}$ in the interaction. 
        \item Reflection symmetry $\mathcal{M}_z$ along the $z$ direction (this preserves the external magnetic field, which is a pseudovector/2-form)
        \begin{equation}
            \psi_{\alpha, k_y, q_z} \mapsto \psi_{-{\alpha}, k_y, -q_z}
        \end{equation}
         This symmetry implies that $h(\{ k_y \})$ in \eqnref{eqn:GeneralInteraction} must satisfy
        \begin{equation}
            h(k_{y,1},k_{y,2},k_{y,3},k_{y,4}) = h(k_{y,3},k_{y,4},k_{y,1},k_{y,2}) .
        \end{equation}
        Using the magnetic translation symmetries, this is equivalent to $h(k_{y,1}-k_{y,2}, k_{y,4}-k_{y,3}) = h(k_{y,2}-k_{y,1}, k_{y,3}-k_{y,4})$, i.~e. $h$ is an even function of its arguments.
        \item $\mathbb{Z}_2$ product of time reversal and reflection along the $x$ direction ($yz$ plane).  Note that this is an antilinear operator that satisfies  $(\mathcal{M}_x\mathcal{T})^2 = 1$. 
        \begin{equation}
            \psi_{\alpha, k_y, q_z} \mapsto \psi_{-{\alpha}, -k_y, -q_z}
        \end{equation}
        This symmetry implies that $h(\{ k_y \})$ in \eqnref{eqn:GeneralInteraction} must satisfy
        \begin{equation}
            h(k_{y,1},k_{y,2},k_{y,3},k_{y,4}) = h(-k_{y,2},-k_{y,1},-k_{y,4},-k_{y,3})^\ast .
        \end{equation}
        As hermiticity of the Hamiltonian implies $h(k_{y,1},k_{y,2},k_{y,3},k_{y,4}) = h(k_{y,4},k_{y,3},k_{y,2},k_{y,1})^\ast$, we see that this is equivalent to
        \begin{equation}
            h(k_{y,1},k_{y,2},k_{y,3},k_{y,4}) = h(-k_{y,3},-k_{y,4},-k_{y,1},-k_{y,2}) .
        \end{equation}
        Using the magnetic translation symmetries, this is equivalent to $h(k_{y,1}-k_{y,2}, k_{y,4}-k_{y,3}) = h(k_{y,4}-k_{y,3}, k_{y,1}-k_{y,2})$, i.~e. $h$ is a symmetric function of its arguments.
    \end{enumerate}
    \item $U(1)$ particle number conservation.
    \begin{equation}
            \psi_{\alpha, k_y, q_z} \mapsto e^{i\phi } \psi_{\alpha, k_y, q_z}  
    \end{equation}
    The interaction is manifestly invariant under this symmetry as we only consider terms with equal number of $\psi$ and $\psi^\dagger$ operators.
    \item $U(1)$ chiral symmetry: conservation of $n_R - n_L$. We neglect Umklapp processes and assume no background electric field, in which case this is an emanent symmetry in the low-energy limit.   
    \begin{equation}
        \psi_{\alpha, k_y, q_z} \mapsto e^{i\alpha \phi} \psi_{\alpha, k_y, q_z}  
    \end{equation}
    
\end{itemize}

\section{Calculation of density profile of nematic CDW}
\label{app:tCDW}
In this appendix, we perform the straightforward calculation of the density profile of the nematic CDW state described in \secref{sec:nCDW}, and shows that it is indeed a charge density wave with a tilted wavevector that breaks rotation symmetry in the plane. The mean field Hamiltonian is given by 
\begin{align}
H_{\text{MF}} &= \int_{q_z,k_y} \term{\psi^\dagger_{R,{q_z,k_y}} \comm{v_F q_z}\, \psi_{R,{q_z,k_yn}} + \psi^\dagger_{L,{q_z,k_y}} \comm{- v_F q_z}\, \psi_{L,{q_z,k_y}} } \\
&\quad + \int_{q_z,k_y} \term{\bar{\Delta} e^{- i k_y k_c \cos\theta }\psi^\dagger_{L,k_y - \frac{k_c \sin\theta}{2}, q_z} \, \psi_{R,k_y + \frac{k_c \sin\theta}{2}, q_z} + \text{h.c.}}
\end{align}
Computing the density 
\begin{equation}
    \expect{\Psi^\dagger(\rr) \Psi(\rr)} \sim \expect{\term{\psi^\dagger_R(\rr)\psi_R(\rr) + \psi^\dagger_L(\rr) \psi_L(\rr)} + \term{e^{2ik_F z} \psi^\dagger_R(\rr)\psi_L(\rr) + \text{h.c.}}}.
    \label{eqn:density_expanded}
\end{equation}
Recall that (in Landau gauge)
\begin{equation}
    \psi_{(R/L)}(\rr) = \int_{k_y, q_z} \frac{1}{(\pi l_B^2)^{1/4}} e^{\term{{i k_y y - (x- k_y l_B^2)^2/2}} } e^{\pm i k_F z + i q z} \psi_{(R/L),k_y,q_z}
\end{equation}

\subsection{Fermion bilinear expectation values}
Note that the right moving fermion $\psi_{R,k_y,q_z,i\omega_n}$ is only coupled to the left moving fermion $\psi_{L,k_y-k_c \sin\theta,q_z,i\omega_n}$ by a coupling $\bar{\Delta}e^{-i\term{k_y -( k_c\sin\theta)/2}k_c\cos\theta}$. This becomes a two-by-two matrix which is easy to diagonalize, and we find (ignoring delta function factors)
\begin{align}
    \expect{\psi^\dagger_{R,k_y,q_z}\psi_{R,k_y,q_z}} &= \half\term{1- \frac{v_Fq_z}{\sqrt{\term{v_Fq_z}^2 + \abs{\Delta}^2}}} \\
    \expect{\psi^\dagger_{L,k_y,q_z}\psi_{L,k_y,q_z}} &= \half\term{1+ \frac{v_Fq_z}{\sqrt{\term{v_Fq_z}^2 + \abs{\Delta}^2}}}
\end{align}
Therefore the first two terms in the expectation value of particle density (\eqnref{eqn:density_expanded}) only contribute a constant uniform background. The crucial part comes from the cross term
\begin{align}
    \expect{\psi^\dagger_{L,k_y - k_c\sin\theta,q_z}\psi_{R,k_y,q_z}} &= \half {\frac{{\Delta}e^{+i\term{k_y -( k_c\sin\theta)/2}k_c\cos\theta}}{\sqrt{\term{v_Fq_z}^2 + \abs{\Delta}^2}}} \\
    \expect{\psi^\dagger_{R,k_y,q_z}\psi_{L,k_y - k_c\sin\theta,q_z}} &= \half {\frac{\bar{\Delta}e^{-i\term{k_y -( k_c\sin\theta)/2}k_c\cos\theta}}{\sqrt{\term{v_Fq_z}^2 + \abs{\Delta}^2}}} 
\end{align}

\subsection{Evaluating the cross term}
The cross term is given by 
\begin{align}
e^{2ik_Fz}\psi_R^\dagger(\mathbf{r},\tau)\psi_L(\mathbf{r},\tau)
&= e^{2ik_Fz}\Biggl[
\int_{dk_y,dq_z}\frac{1}{(\pi l_B^2)^{1/4}}
\exp\Bigl[-ik_y y-\frac{(x-k_yl_B^2)^2}{2l_B^2}\Bigr]
\exp\Bigl[-iq_z z\Bigr]
\psi_{R,k_y,q_z}^\dagger
\Biggr]\nonumber\\[1mm]
&\quad\times\Biggl[
\int_{dk'_y ,dq'_z }\frac{1}{(\pi l_B^2)^{1/4}}
\exp\Bigl[ik'_yy-\frac{(x-k'_yl_B^2)^2}{2l_B^2}\Bigr]
\exp\Bigl[iq'_z z\Bigr]
\psi_{L,k'_y,q'_z}
\Biggr].
\label{eq:psiRdag-psiL-step}
\end{align}
We know the expectation value is non-zero if and only if $q_z=q_z'$ and $k_y' = k_y - k_c\sin\theta$. Therefore, the expectation value becomes (we add factors of $l_B$ here explicitly for clarity),
\begin{align}
e^{2ik_Fz}\expect{\psi_R^\dagger(\mathbf{r},\tau)\psi_L(\mathbf{r},\tau)}
&= e^{2ik_Fz}
\int_{dk_y,dq_z}\Biggl[\frac{1}{(\pi l_B^2)^{1/2}}
\exp\Bigl[-ik_y y-\frac{(x-k_yl_B^2)^2}{2l_B^2}\Bigr]
\exp\Bigl[-iq_z z\Bigr]
\Biggr]\nonumber\\[1mm]
&\times\Biggl[
\exp\Bigl[i\term{k_y-k_c\sin\theta}y-\frac{\term{x-\term{k_y-k_c\sin\theta}l_B^2}^2}{2l_B^2}\Bigr]
\exp\Bigl[iq_z z\Bigr]
\Biggr]
\expect{\psi_{R,k_y,q_z}^\dagger\psi_{L,\term{k_y-k_c\sin\theta},q_z}},\\
&= e^{2ik_Fz}\Biggl[
\int_{dk_y,dq_z}\frac{1}{(\pi l_B^2)^{1/2}}
\exp\Bigl[-ik_c\sin\theta  y-\frac{(x-k_yl_B^2)^2}{2l_B^2}-\frac{\term{x-\term{k_y-k_c\sin\theta}l_B^2}^2}{2l_B^2}\Bigr]
\nonumber\\[1mm]
&\quad \times
\expect{\psi_{R,k_y,q_z}^\dagger\psi_{L,\term{k_y-k_c\sin\theta},q_z}}\Biggr],\\
&= e^{2ik_Fz}\Biggl[
\int_{dk_y,dq_z}\frac{1}{(\pi l_B^2)^{1/2}}
\exp\Bigl[-ik_c\sin\theta  y-\frac{(x-k_yl_B^2)^2}{2l_B^2}-\frac{\term{x-\term{k_y-k_c\sin\theta}l_B^2}^2}{2l_B^2}\Bigr]
\nonumber\\[1mm]
&\quad \times
\half {\frac{\bar{\Delta}e^{-il_B^2\term{k_y -( k_c\sin\theta)/2}k_c\cos\theta}}{\sqrt{\term{v_Fq_z}^2 + \abs{\Delta}^2}}}\Biggr].
\label{eq:psiRdag-psiL-step2}
\end{align}
The integral over $q_z$ just contributes a cutoff dependent factor. We can write
\begin{align}
e^{2ik_Fz}\expect{\psi_R^\dagger(\mathbf{r},\tau)\psi_L(\mathbf{r},\tau)}
&\sim e^{2ik_Fz}\Biggl[
\int_{dk_y}\frac{1}{(\pi l_B^2)^{1/2}}
\exp\Bigl[-ik_c\sin\theta  y-\frac{(x-k_yl_B^2)^2}{2l_B^2}-\frac{\term{x-\term{k_y-k_c\sin\theta}l_B^2}^2}{2l_B^2}\Bigr]
\nonumber\\[1mm]
&\quad \times
{{\bar{\Delta}\exp\comm{-il_B^2\term{k_y -( k_c\sin\theta)/2}k_c\cos\theta}}}\Biggr].
\label{eq:psiRdag-psiL-step3}
\end{align}
Now, we have a Gaussian integral for $k_y$, performing this integral, we find
\begin{align}
    e^{2ik_Fz}\expect{\psi_R^\dagger(\mathbf{r},\tau)\psi_L(\mathbf{r},\tau)} \sim\frac{\pi^2 \bar{\Delta} }{2} e^{-k_c^2 l_B^2/4} e^{2ik_Fz-i\term{ k_c\cos(\theta)x + k_c\sin(\theta)y} }
\end{align}
Adding the complex conjugate, we find the total density expectation value (we write $\Delta = \abs{\Delta}e^{i\phi_0}$)
\begin{equation}
    \expect{\Psi^\dagger(\rr) \Psi(\rr)} = \rho_0 + A \abs{\Delta} \cos\term{2k_F z - k_c\cos(\theta)x - k_c\sin(\theta)y - \phi_0} 
    \label{eqn:density_final}
\end{equation}
where $A$ is some constant factor. Therefore, we have shown that the nematic CDW mean field Hamiltonian has a density modulation with wavefector $\kk = (-\kk_c, 2k_F)$.

\section{BCS mean field theory for superconducting layer}
\label{app:BCSMF}
Consider a single superconducting island with the effective truncated Hamiltonian $H = H_0 + H_\text{int}$ where
\begin{equation}
    H_0 = \sum_{k_y, k_z} \term{\frac{k_z^2}{2m}  - V_0 e^{-\pi^2 l_B^2 / \lambda^2} \cos\term{\frac{2 \pi k_y l_B^2}{\lambda}} - \mu} c^\dagger_{\kk} c_{\kk} \approx \sum_{k_y, q_z; \alpha = \pm 1} \term{\alpha v_F q_z} \psi^\dagger_{\alpha, k_y, q_z} \psi_{\alpha, k_y, q_z},
\end{equation}
where $q_z$ is the $z$ momentum relative to the Fermi surface $k_z(k_y)$. The interactions come from projecting a contact interaction between the right and left movers into the LLL, and then truncating to keep only the BCS interactions. 
\begin{equation}
    H_\text{int} = -g\sum_{k_y, k_y';-\hbar\omega_D < q_z, q_z' < \hbar\omega_D } \psi^\dagger_{+, k_y, q_z} \psi^\dagger_{-, -k_y, -q_z} \psi_{-, -k_y', -q_z'} \psi_{+, k_y', q_z'} e^{-2 l_B^2(k_y^2 + k_y'^2)}.
\end{equation}
We mean-field decompose this interaction in the BCS channel and self-consistently solve the gap equation. To do so, we define ($\qq \equiv (k_y, q_z)$)
\begin{align}
    d_{\qq} &= \expect{\psi_{-,-\qq} \psi_{+,+\qq}} \\
    \Delta_{\qq} &= -g \sum_{\qq'} d_{\qq'} e^{-2l_B^2(k_y^2 + k_y'^2)} = -g e^{-2l_B^2 k_y^2 } \sum_{\qq'} d_{\qq'} e^{-2l_B^2 k_y'^2}
\end{align}
Thus we have the mean-field BCS Hamiltonian
\begin{equation}
    H_\text{BCS} = \sum_{\qq} v_F q_z\term{\psi^\dagger_{+,\qq}\psi_{+,\qq} - \psi^\dagger_{-,-\qq} \psi_{-,-\qq}} + \Delta_{\qq}^\ast \psi_{-,-\qq} \psi_{+,\qq} + \Delta_{\qq} \psi^\dagger_{+,\qq} \psi^\dagger_{-,-\qq}.
\end{equation}
This is in the standard Boguiliubov-de Gennes (BdG) form and we can diagonalize it by defining 
\begin{equation}
    \gamma_{\qq} = \begin{pmatrix} \psi_{+,+\qq} \\ \psi^\dagger_{-,-\qq} \end{pmatrix} 
\end{equation}
and the BCS mean-field Hamiltonian is $H = \sum_{\qq} \gamma_{\qq}^\dagger \mathcal{H}_{\qq} \gamma_{\qq}$, where
\begin{equation}
    \mathcal{H}_{\qq} = \half \begin{pmatrix} v_F q_z & \Delta_{\qq}^\ast \\ \Delta_{\qq} & -v_F q_z\end{pmatrix}
\end{equation}
Using the standard BdG machinery, we diagonalize this using a unitary matrix
\begin{equation}
    U_{\qq} = \begin{pmatrix} \cos \theta_{\qq} & \sin \theta_{\qq} \\ \sin \theta_{\qq} & -\cos \theta_{\qq} \end{pmatrix}, \tan 2\theta_{\qq} = -\Delta_{\qq} / (v_F q_z), \qquad U_{\qq} \mathcal{H}_{qq} U_{\qq}^\dagger = \half \begin{pmatrix} \lambda_{\qq} & 0 \\ 0 & -\lambda_{\qq}\end{pmatrix}, \lambda_{\qq} = \sqrt{v_F^2 q_z^2 + \abs{\Delta_{\qq}}^2}
\end{equation}
It is also useful to define the Boguliubov quasiparticle operator 
\begin{equation}
    \begin{pmatrix} \chi_{\qq,1} \\ \chi_{\qq,2} \end{pmatrix} = U_{\qq}  \begin{pmatrix} \psi_{+,+\qq} \\ \psi^\dagger_{-,-\qq} \end{pmatrix}
\end{equation}
having zero off-diagonal expectation values and non-zero diagonal expectation values (at finite temperature)
\begin{equation}
    \expect{\chi^\dagger_{\qq,1} \chi_{\qq,1}} = n_F(\lambda_{\qq}) \qquad \expect{\chi^\dagger_{\qq,2} \chi_{\qq,2}} = 1- n_F(\lambda_{\qq})
\end{equation}
where $n_F$ is the Fermi-Dirac distribution function. Using these relations, we can find the self-consistent gap equation at finite temperature
\begin{equation}
    \Delta_{\qq} = g e^{-2 l_B^2 q_y^2} \sum_{\qq'} \frac{\Delta_{\qq'}}{2\lambda_{\qq'}} \tanh\term{\beta \lambda_{\qq'}/2} e^{-2 l_B^2 q_y'^2} 
\end{equation}
we can simplify this by using the motivated ansatz $\Delta_{\qq} = \Delta_0 e^{-2 l_B^2 q_y^2}$. Plugging this is, we get
\begin{equation}
    1 = \frac{g}{2} \int_{-\Lambda}^{\Lambda} \frac{dq_z}{2\pi} \int_{-\infty}^\infty \frac{dq_y}{2\pi}  \frac{e^{-4 l_B^2 q_y^2} }{\lambda_{\qq}} \tanh\term{\beta \lambda_{\qq}/2} 
\end{equation}
We can solve this to get the gap-to-$T_c$ ratio. The zero temperature gap is given by setting $\beta \to \infty$.
\begin{equation}
    1 = \frac{g}{2} \int_{-\Lambda}^{\Lambda} \frac{dq_z}{2\pi} \int_{-\infty}^\infty \frac{dq_y}{2\pi}  \frac{e^{-4 l_B^2 q_y^2} }{\sqrt{v_F^2 q_z^2 + \abs{\Delta_{0}(T=0)}^2 e^{-4l_B^2 q_y^2}}} 
\end{equation}
Similarly, the critical temperature can be calculated by setting $\Delta_0 \to 0$. 
\begin{equation}
    1 = \frac{g}{2} \int_{-\Lambda}^{\Lambda} \frac{dq_z}{2\pi} \int_{-\infty}^\infty \frac{dq_y}{2\pi}  \frac{e^{-4 l_B^2 q_y^2} }{v_F \abs{q_z}} \tanh\term{\beta_c v_F \abs{q_z}/2} 
\end{equation}
The critical temperature can be evaluated approximately by using 
\begin{equation}
    \frac{\tanh(x)}{x} \lesssim \begin{cases} 1 & \abs{x}\leq 1 \\ \frac{1}{\abs{x}} & \abs{x}>1\end{cases}
\end{equation}
To find 
\begin{equation}
    k_B T_c \lesssim \frac{\exp(1)}{2} \hbar \omega_D  e^{-\frac{2\sqrt{2}}{g(2\pi v_F \sqrt{2 \pi l_B^2})^{-1}}} 
\end{equation}
where the coefficient in the exponent plays the role of an effective density of states at the Fermi surface. We can also calculate the gap-to-$T_c$ ratio numerically without making any drastic approximations by substracting the two equations and sending $\omega_D \to \infty$. This gives us 
\begin{equation}
    \frac{\Delta_0(T=0)}{k_BT_c}\approx 2.26
\end{equation}

\section{Calculation of the condensate order parameter profile}
\label{app:CooperProfile}
In this appendix, we compute the condensate order parameter profile $\Psi_{\text{GL}}(\rr) = \expect{\psi_L(\rr)\psi_R(\rr)}$ in the bulk superconducting phase described in \secref{sec:Bulk_Superconductor}. We work in units where $l_B = 1$ for simplifying notation. The mean field Hamiltonian is given by
\begin{equation}
    \mathcal{H}_{\qq} = \begin{pmatrix}
        \psi^\dagger_{R,\qq} \\ \psi_{L,-\qq}
    \end{pmatrix}^T
    \begin{pmatrix}
        v_F q_z & \tilde \Delta(q_x,q_y) \\
        \tilde \Delta^\ast(q_x,q_y) & -v_F q_z
    \end{pmatrix}
    \begin{pmatrix}
        \psi_{R,\qq} \\
        \psi^\dagger_{L,-\qq}
    \end{pmatrix}
\end{equation}
To calculate the condensate order parameter profile, we need to calculate
\begin{equation}
    \Psi_{\text{GL}}(\rr) = \expect{\psi_L(\rr)\psi_R(\rr)} = \int_{k_y, q_z} \int_{k_y', q_z'} \frac{e^{i(k_y + k_y')y}}{\pi^{1/2}} e^{-\frac{(x-k_y)^2}{2}} e^{-\frac{(x-k_y')^2}{2}} e^{i(q_z + q_z')z} \expect{\psi_{L,k_y', q_z'} \psi_{R,k_y, q_z}}
\end{equation}
From the Bogoliubov Hamiltonian, we know that
\begin{equation}
    \expect{\psi_{L,\qq} \psi_{R,-\qq}} = \frac{\tilde \Delta(q_x,q_y)}{2 \sqrt{v_F^2 q_z^2 + |\tilde \Delta(q_x,q_y)|^2}},
\end{equation} 
which means that 
\begin{align}
    \expect{\psi_{L,K_y' + q_y',q_z'} \psi_{R,K_y + q_y, q_z}} = \int_{q_x, q_x'} e^{-i K_y q_x - i K_y' q_x'} \expect{\psi_{L,\qq'} \psi_{R,\qq}} =\int_{q_x} e^{-i (K_y - K_y') q_x}  \frac{\tilde \Delta(q_x,q_y)}{2 \sqrt{v_F^2 q_z^2 + |\tilde \Delta(q_x,q_y)|^2}}.
\end{align} 
Plugging this back in, we get
\begin{align}
\Psi_{\text{GL}}(\rr) &=  \sum_{K_y, K_y'}\int_{q_y, q_z}  \frac{e^{i(K_y + K_y')y}}{\pi^{1/2}} e^{-\frac{(x-(K_y+q_y))^2}{2}} e^{-\frac{(x-(K_y' - q_y)))^2}{2}} \comm{\int_{q_x} e^{-i (K_y - K_y') q_x}  \frac{\tilde \Delta(q_x,q_y)}{2 \sqrt{v_F^2 q_z^2 + |\tilde \Delta(q_x,q_y)|^2}}}   \\
&=  \sum_{K_y, K_y'}\int_{q_x, q_y, q_z}  \frac{e^{i(K_y + K_y')y}}{\pi^{1/2}} e^{-\term{x-\frac{(K_y+K_y')}{2}}^2 -\term{\frac{K_y - K_y'}{2} - q_y}^2} {e^{-i (K_y - K_y') q_x}  \frac{\tilde \Delta(q_x,q_y)}{2 \sqrt{v_F^2 q_z^2 + |\tilde \Delta(q_x,q_y)|^2}}} 
\end{align}
Recall that $K_y, K_y' \in \lambda \mathbb{Z}/2 $. Instead of trying to simplify this expression, we will try to identify the zeros of the condensate order parameter using its symmetries. First, we identify a symmetry in the expression for $\tilde \Delta(q_x,q_y)$
\begin{equation}
    \tilde \Delta(q_x,q_y) = \sum_{K_y} e^{-i 2 K_y q_x} \Delta_0(K_y+q_y) = e^{-i q_x \lambda}\tilde \Delta(-q_x,\lambda / 2-q_y) = \tilde \Delta(q_x+ 2 \pi  \lambda^{-1}, q_y) .
\end{equation}
 Let us focus on the point $x = \lambda / 4, y =  \pi \lambda^{-1}$. In the integrand (summand), we perform the following transformation that keeps the integral (sum) invariant
\begin{equation}
    K_y \to -\lambda / 2 + K_y' , \quad  K_y' \to \lambda / 2 + K_y, \quad \qq \to (-q_x, \lambda / 2 - q_y, -q_z)      
     \label{eqn:tfmn1}              
\end{equation}
After this transformation, we have
\begin{align}
\Psi_{\text{GL}}(\rr) &=  \sum_{K_y, K_y'}\int_{q_x, q_y, q_z}  \frac{e^{i \pi - i(K_y + K_y')\pi \lambda^{-1}}}{\pi^{1/2}} e^{-\term{ - \lambda /2 + \frac{(K_y+K_y')}{2}}^2 - \term{\frac{K_y - K_y'}{2} + q_y}^2} {e^{-i (K_y - K_y') q_x}  \frac{\tilde \Delta(q_x,q_y)}{2 \sqrt{v_F^2 q_z^2 + |\tilde \Delta(q_x,q_y)|^2}}}  
\end{align}
To understand why the above sum is zero, it is useful to divide $\sum_{K_y,K_y'}$ into two parts, one where $K_y + K_y' = \lambda n$ for $n \in \mathbb{Z}$, and the other where $K_y + K_y' = \lambda (n + 1/2)$. In the first (even) case, the summand is simply the negative of the original summand, and therefore that sum is zero. In the second (odd) case, one can use the symmetry $\tilde \Delta(q_x, q_y) = \tilde \Delta(q_x + 2\pi \lambda^{-1},q_y)$ to write
\begin{align}
    &\sum_{\substack{K_y, K_y',\\ K_y + K_y' \in \frac{\lambda}{2}\term{\mathbb{Z}+\half} }}\int_{q_x = -2\pi/\lambda}^0 \int_{q_y, q_z}  \frac{e^{i \pi - i(K_y + K_y')\pi \lambda^{-1}}}{\pi^{1/2}} e^{-\term{ - \lambda /2 + \frac{(K_y+K_y')}{2}}^2 - \term{\frac{K_y - K_y'}{2} + q_y}^2} \nonumber \\
    &\qquad \qquad \times \term{e^{-i (K_y - K_y') q_x} + e^{-i (K_y - K_y') (q_x + 2\pi/\lambda)}}  \frac{\tilde \Delta(q_x,q_y)}{2 \sqrt{v_F^2 q_z^2 + |\tilde \Delta(q_x,q_y)|^2}}  
\end{align}
which is zero as the term in the bottom paranthesis is zero when $K_y - K_y'$ is an off multiple of $\lambda/2$.

These symmetry arguments can be used to show that $\Psi_{\text{GL}}(x,y) = e^{-i\lambda y/l_B^2} \Psi_{\text{GL}}(x+\lambda/2, y) = \Psi_{\text{GL}}(x, y+2\pi l_B^2/\lambda)$ (note that we were working in units where $l_B = 1$).
Thus, we have shown that the condensate order parameter profile has a rectangular lattice of zeros at these points, where the area of the unit cell is $\pi l_B^2$. This is consistent with the fact that each vortex carries a flux quantum $h/2e$, giving us two flux quanta per magnetic unit cell.

\section{Mean-field theory of the triangular vortex lattice}
\label{app:tr_vortex_lattice}
In the main text, we discussed the case of the mean-field pairing function $\Delta(K_y,k_y) = \Delta_0(k_y)$ and argued that this produces a rectangular vortex lattice, with two Weyl nodes in the Boguliubov spectrum. In this section, we show that 
\begin{equation}
    \Delta(K_y,k_y) = (-1)^{\half n_{K_y} \term{n_{K_y} + 1}}\Delta_0(k_y),\qquad n_{K_y} \equiv \frac{K_y l_B^2}{\lambda / 2},
\end{equation} 
produces a triangular vortex lattice and study its quasiparticle spectrum. We have the mean-field Hamiltonian $H = H_{\mathrm{KE}} + H_{\mathrm{Pair}}$, where
\begin{align}
    H_{\text{KE}} &= \int_{k_y,q_z} v_F q_z\term{\psi^\dagger_{R, k_y,q_z}\psi_{R,k_y,q_z} - \psi^\dagger_{L,k_y,q_z}\psi_{L,k_y,q_z}}, 
    \label{eq:HKE_tri} \\
    H_\text{Pair} &= \sum_{K_y \in \frac{\lambda}{2 l_B^2} \mathbb{Z}} \int_{k_y,q_z} \term{\Delta(K_y, k_y) \psi^\dagger_{R,K_y + k_y,q_z} \psi^\dagger_{L,K_y-k_y,-q_z} + \text{h.c.}}, \\
    &= \sum_{K_y \in \frac{\lambda}{2 l_B^2} \mathbb{Z}} e^{i \frac{\pi}{2} n_{K_y}(n_{K_y} + 1)} \int_{k_y,q_z} \term{\Delta_0(k_y) \psi^\dagger_{R,K_y + k_y,q_z} \psi^\dagger_{L,K_y-k_y,-q_z} + \text{h.c.}}
    \label{eq:Hpair_tri}
\end{align}
To motivate the correct basis to diagonalize this Hamiltonian, we rewrite
\begin{align}
    H_{\text{KE}} &= \sum_{K'_y  \in \frac{\lambda}{2 l_B^2} \mathbb{Z}} \int_{q_z} \int_0^{\lambda/(2l_B^2)} \frac{dq_y}{2\pi} v_F q_z\term{\psi^\dagger_{R, K'_y + q_y,q_z}\psi_{R,K'_y+q_y,q_z} - \psi_{L,-K'_y - q_y ,-q_z}\psi^\dagger_{L,-K'_y - q_y,-q_z}}, 
    \label{eq:HKE_tri2} \\
    H_\text{Pair} &= \sum_{K_y,K'_y \in \frac{\lambda}{2 l_B^2} \mathbb{Z}} e^{i \frac{\pi}{2} n_{K_y}(n_{K_y} + 1)} \int_{q_z} \int_0^{\lambda/(2l_B^2)}  \frac{dq_y}{2\pi}  \term{\Delta_0(K'_y + q_y) \psi^\dagger_{R,K_y +  K'_y + q_y,q_z} \psi^\dagger_{L,K_y- K'_y - q_y,-q_z} + \text{h.c.}}.
    \label{eq:Hpair_tri2}
\end{align}
This motivates the Fourier transformed fields
\begin{align}
    \psi_{R, K_y +q_y, q_z} &= \int_{-2\pi/\lambda}^{2\pi/\lambda} \frac{dq_x}{4\pi / \lambda} e^{-i q_x K_y l_B^2} e^{-i \frac{\pi}{4} n_{K_y}(n_{K_y} + 1)} \psi_{R,q_x,q_y,q_z}, \\
    \psi_{L, K_y - q_y, -q_z} &= \int_{-2\pi/\lambda}^{2\pi/\lambda} \frac{dq_x}{4\pi / \lambda} e^{i q_x K_y l_B^2} e^{-i \frac{\pi}{4} n_{K_y}(n_{K_y} + 1)} \psi_{L,-q_x,-q_y,-q_z}.
\end{align}
where $q_y$ is restricted to the first magnetic Brillouin zone which we define to be $[0, \lambda/(2l_B^2))$. Plugging this in to the Hamiltonian, we find
\begin{align}
    H_{\text{KE}} &= \sum_{K'_y  \in \frac{\lambda}{2 l_B^2} \mathbb{Z}} \int_{q_z} \int \frac{dq_y}{2\pi} v_F q_z\int \frac{dq_x dq'_x}{(4\pi/\lambda)^2}e^{i(-q_x +q'_x)K'_y l_B^2}\term{\psi^\dagger_{R, q_x, q_y,q_z}\psi_{R,q'_x , q_y,q_z}  - \psi_{L,-q_x, - q_y ,-q_z}\psi^\dagger_{L,-q'_x, - q_y,-q_z}}, \\
    &= \int \frac{dq_x dq_y dq_z}{(2\pi)^2(4\pi/\lambda)} v_F q_z\term{\psi^\dagger_{R, q_x, q_y,q_z}\psi_{R,q_x , q_y,q_z}  - \psi_{L,-q_x, - q_y ,-q_z}\psi^\dagger_{L,-q_x,- q_y,-q_z}},  
    \label{eq:HKE_tri3} \\
    H_\text{Pair} &= \sum_{K_y,K'_y \in \frac{\lambda}{2 l_B^2} \mathbb{Z}} e^{i \frac{\pi}{4} \term{2 n_{K_y}(n_{K_y} + 1) + (n_{K_y}+n_{K'_y})(n_{K_y}+n_{K'_y}+1) + (n_{K_y}-n_{K'_y})(n_{K_y}-n_{K'_y}+1)}} \nonumber \\
    &\qquad \times \int_{q_z} \int  \frac{dq_y dq_x dq'_x}{2\pi (4\pi/\lambda)^2}  \term{\Delta_0(K'_y + q_y) \psi^\dagger_{R,q_x,q_y,q_z} \psi^\dagger_{L,-q'_x, - q_y,-q_z} e^{iq_x(K_y + K_y')l_B^2  - i q'_x(K_y - K_y')l_B^2}+ \text{h.c.}}, \\
    &= \sum_{K'_y \in \frac{\lambda}{2 l_B^2} \mathbb{Z}} e^{i \frac{\pi}{4} \term{ 2n_{K_y'}^2}} \int_{q_z} \int  \frac{dq_y dq_x }{2\pi (4\pi/\lambda)}  \term{\Delta_0(K'_y + q_y) \psi^\dagger_{R,q_x,q_y,q_z} \psi^\dagger_{L,-q_x, - q_y,-q_z} e^{-i2 q_x K_y' l_B^2 }+ \text{h.c.}}, \\
    &=   \int  \frac{ dq_x dq_y dq_z}{(2\pi)^2 (4\pi/\lambda)}  \comm{\term{\sum_{K'_y \in \frac{\lambda}{2 l_B^2} \mathbb{Z}} e^{i \frac{\pi}{2} n_{K_y'}^2} e^{-i2 q_x K_y' l_B^2 } \Delta_0(K'_y + q_y)} \psi^\dagger_{R,q_x,q_y,q_z} \psi^\dagger_{L,-q_x, - q_y,-q_z} + \text{h.c.}}, \\
    &\equiv   \int  \frac{ dq_x dq_y dq_z}{(2\pi)^2 (4\pi/\lambda)}  \comm{\tilde \Delta_0'(q_x,q_y) \psi^\dagger_{R,q_x,q_y,q_z} \psi^\dagger_{L,-q_x, - q_y,-q_z} + \text{h.c.}} ,
    \label{eq:Hpair_tri3}
\end{align}
which is of the standard BdG form. We can diagonalize this Hamiltonian by using the usual Boguliubov transformation, and find the quasiparticle spectrum. The quasiparticle spectrum has nodes when $\tilde \Delta_0'(q_x,q_y) = 0$ and $q_z = 0$. Examining the expression
\begin{equation}
    \tilde \Delta_0'(q_x,q_y) = \sum_{K'_y \in \frac{\lambda}{2 l_B^2} \mathbb{Z}} e^{i \frac{\pi}{2} n_{K_y'}^2} e^{-i2 q_x K_y' l_B^2 } \Delta_0(K'_y + q_y)
\end{equation}
and recalling that $\Delta_0(K'_y + q_y)$ is an even function, we see that $\tilde \Delta_0'(q_x,q_y)$ is zero when $q_y = \lambda/(4 l_B^2)$ and $e^{i \pi / 2 + i 2 q_x \lambda /2} = -1 \implies q_x = \pi / (2 \lambda)$ or $q_x = -3 \pi / (2\lambda)$. This is again consistent with the fact that the vortex lattice has two vortices per magnetic unit cell, and therefore we expect two Weyl nodes in the quasiparticle spectrum.

To see the triangular vortex lattice structure in real space, we can calculate the condensate order parameter $\Psi_{\text{GL}}(\rr) = \expect{\psi_L(\rr)\psi_R(\rr)}$ using the same method as in \appref{app:CooperProfile}. To structure the analysis, we first calculate the local particle creation operator in terms of the Fourier transformed fields $\psi_{R,q_x,q_y,q_z}$ and $\psi_{L,q_x,q_y,q_z}$. We work in units of $l_B = 1$ to simplify notation. We have
\begin{align}
\psi_R(\rr) &= \int_{k_y, q_z} \frac{e^{i k_y y}}{\pi^{1/4}} e^{-\frac{(x-k_y)^2}{2}} e^{i q_z z} \psi_{R,k_y,q_z}, \\
&= \sum_{K_y} \int_{q_y, q_z} \frac{e^{i k_y y}}{\pi^{1/4}} e^{-\frac{(x-K_y - q_y)^2}{2}} e^{i q_z z} \int_{-2\pi/\lambda}^{2\pi/\lambda} \frac{dq_x}{4\pi/\lambda} e^{-i q_x K_y} e^{-i \frac{\pi}{4} n_{K_y} (n_{K_y} + 1)} \psi_{R,q_x,q_y,q_z}, \\
&= \int_{q_x, q_y, q_z} \frac{e^{i (K_y + q_y) y}}{\pi^{1/4}}  e^{i q_z z} \comm{\sum_{K_y} e^{-i q_x K_y} e^{-i \frac{\pi}{4} n_{K_y} (n_{K_y} + 1)}e^{-\frac{(x-K_y - q_y)^2}{2}}} \psi_{R,q_x,q_y,q_z}.
\end{align}
Recall that $K_y$ is quantized in units of $\lambda /2 $. Similarly, we have
\begin{align}
\psi_L(\rr) &= \int_{q_x, q_y, q_z} \frac{e^{i (K_y - q_y) y}}{\pi^{1/4}}  e^{-i q_z z} \comm{\sum_{K_y} e^{+i q_x K_y} e^{-i \frac{\pi}{4} n_{K_y} (n_{K_y} + 1)} e^{-\frac{(x-K_y + q_y)^2}{2}}} \psi_{R,-q_x,-q_y,-q_z}.
\end{align}
Putting this together, we can easily write down an expression for $\expect{\psi_L(\rr)\psi_R(\rr)}$ in terms of $\expect{\psi_{L,\qq} \psi_{R,-\qq'}} = \tilde \Delta'(\qq)/(v_F^2 q_z^2 + \vert{\tilde \Delta'(\qq)}\vert^2)^{1/2} \delta^{(3)}(\qq - \qq')$ which we know from the BdG Hamiltonian. To avoid writing down a cumbersome expression, we skip the step of integrating over $\qq'$ by using the delta function to find
\begin{align}
\expect{\psi_L(\rr)\psi_R(\rr)} &\propto  \int_{q_x, q_y, q_z}  \comm{\sum_{K_y} e^{i K_y y} e^{-i q_x K_y} e^{-i \frac{\pi}{4} n_{K_y} (n_{K_y} + 1)}e^{-\frac{(x-K_y - q_y)^2}{2}}} \comm{\sum_{K'_y} e^{i K_y' y} e^{i q_x K'_y} e^{-i \frac{\pi}{4} n_{K'_y} (n_{K'_y} + 1)} e^{-\frac{(x-K'_y + q_y)^2}{2}}} \nonumber \\
&\qquad \times \frac{\tilde \Delta'(q_x,q_y)}{\sqrt{v_F^2 q_z^2 + \vert{\tilde \Delta'(q_x,q_y)}\vert^2}} , \\
&= \sum_{K_y,K'_y} \int_{q_x, q_y, q_z} \frac{1}{\pi^{1/2}}  e^{i (K_y + K_y') y} e^{-i q_x (K_y - K'_y)} e^{-i \frac{\pi}{4} \term{n_{K_y}(n_{K_y} + 1) + n_{K'_y}(n_{K'_y} + 1)}} \nonumber \\
&\qquad \times e^{-\frac{(x-K_y - q_y)^2}{2} -\frac{(x-K'_y + q_y)^2}{2}} \frac{\tilde \Delta'(q_x,q_y)}{\sqrt{v_F^2 q_z^2 + \vert{\tilde \Delta'(q_x,q_y)}\vert^2}}.
\end{align}
For simplicity of notation, we introduce integers $n, n'$ such that $K_y = n \lambda /2, K_y' = n' \lambda /2$ and we also define $n^\pm = n \pm n'$. Then, we have
\begin{align}
\expect{\psi_L(\rr)\psi_R(\rr)} &\propto \sum_{n,n'} \int_{q_x, q_y, q_z}  e^{i n^+ y \lambda/2} e^{-i q_x n^- \lambda /2 } e^{-i \frac{\pi}{4}  \term{n^+ + (n^+)^2/2 + (n^-)^2/2}} \nonumber \\
&\qquad \times e^{-(x-n^+ \lambda / 4)^2 - \term{q_y - n^- \lambda / 4}^2} \frac{\tilde \Delta'(q_x,q_y)}{\sqrt{v_F^2 q_z^2 + \vert{\tilde \Delta'(q_x,q_y)}\vert^2}}.
\end{align}
Let us go back and examine the symmetries of $\tilde\Delta'$. We note that 
\begin{equation}
    \tilde \Delta'(q_x, q_y) = \tilde \Delta'(q_x + 2\pi/\lambda, q_y) = i e^{-i q_x \lambda} \tilde\Delta'(-q_x - \pi/\lambda , -q_y + \lambda / 2).
\end{equation}
Let us examine the point $x = \lambda /4, y = 0$. The RHS evaluates to 
\begin{align}
\expect{\psi_L(\lambda/4,0)\psi_R(\lambda/4,0)} &\propto \sum_{n,n'} \int_{q_x, q_y, q_z}   e^{-i q_x n^- \lambda /2 } e^{-i \frac{\pi}{4}  \term{n^+ + (n^+)^2/2 + (n^-)^2/2}} \nonumber \\
&\qquad \times e^{-(\lambda / 4 -n^+ \lambda / 4)^2 - \term{q_y - n^- \lambda / 4}^2} \frac{\tilde \Delta'(q_x,q_y)}{\sqrt{v_F^2 q_z^2 + \vert{\tilde \Delta'(q_x,q_y)}\vert^2}}.
\end{align}
Next, we apply the transformation $n \mapsto 1 - n', n\mapsto 1- n$ that takes $n^+ \mapsto 2 - n^+, n^- \mapsto n^-$. Then we have 
\begin{align}
\expect{\psi_L(\lambda/4,0)\psi_R(\lambda/4,0)} &\propto \sum_{n,n'} \int_{q_x, q_y, q_z}   e^{-i q_x n^- \lambda /2 } e^{-i \frac{\pi}{4}  \term{2 - n^+ + 2 - 2 n^+ + (n^+)^2/2 + (n^-)^2/2}} \nonumber \\
&\qquad \times e^{-(\lambda / 4 -n^+ \lambda / 4)^2 - \term{q_y - n^- \lambda / 4}^2} \frac{\tilde \Delta'(q_x,q_y)}{\sqrt{v_F^2 q_z^2 + \vert{\tilde \Delta'(q_x,q_y)}\vert^2}},\\
&= \sum_{n,n'} e^{-i\pi(1-n^+)} \int_{q_x, q_y, q_z}   e^{-i q_x n^- \lambda /2 } e^{-i \frac{\pi}{4}  \term{n^+ + (n^+)^2/2 + (n^-)^2/2}} \nonumber \\
&\qquad \times e^{-(\lambda / 4 -n^+ \lambda / 4)^2 - \term{q_y - n^- \lambda / 4}^2} \frac{\tilde \Delta'(q_x,q_y)}{\sqrt{v_F^2 q_z^2 + \vert{\tilde \Delta'(q_x,q_y)}\vert^2}}.
\end{align}
If $n^+$ is even, the above expression is the negative of the previous expression and therefore zero. When $n^+$ is an odd number, then $n^-$ is also odd and one can write $\int_{q_z} f(q_z) = \int_{q_z = -2\pi/\lambda}^0 (f(q_z)+ f(q_z+ 2\pi/\lambda)$. Using the symmetry of $\tilde \Delta'$ and that $e^{-i(q_x + 2\pi / \lambda) n^-\lambda / 2}= -e^{-i q_x  n^-\lambda / 2}$, we conclude that the sum is zero and that the order parameter is zero at $\rr = (\lambda/4,0)$. 

To identify the lattice structure, we first note that 
\begin{align}
\expect{\psi_L(x, y+2\pi / \lambda)\psi_R(x, y + 2\pi / \lambda)} &\propto \sum_{n,n'} \int_{q_x, q_y, q_z}  e^{i n^+ (y + 2 \pi /\lambda)  \lambda/2} e^{-i q_x n^- \lambda /2 } e^{-i \frac{\pi}{4}  \term{n^+ + (n^+)^2/2 + (n^-)^2/2}} \nonumber \\
&\qquad \times e^{-(x-n^+ \lambda / 4)^2 - \term{q_y - n^- \lambda / 4}^2} \frac{\tilde \Delta'(q_x,q_y)}{\sqrt{v_F^2 q_z^2 + \vert{\tilde \Delta'(q_x,q_y)}\vert^2}}.
\end{align}
We see that this is equal when $n^+$ is even, and the $n^+$ odd sum vanishes because of the same $q_x \to q_x + 2\pi / \lambda$ symmetry discussed earlier. Therefore, we have 
\begin{equation}
    \expect{\psi_L(x, y+2\pi / \lambda)\psi_R(x, y + 2\pi / \lambda)} = \expect{\psi_L(x, y)\psi_R(x, y)},
\end{equation}
with the equality because the proportionality constant we did not keep track of is the same for both expectation values. Next, consider
\begin{align}
\expect{\psi_L(x + \lambda/2, y+\pi / \lambda)\psi_R(x + \lambda / 2, y + \pi / \lambda)} &\propto \sum_{n,n'} \int_{q_x, q_y, q_z}  e^{i n^+ (y +  \pi /\lambda)  \lambda/2} e^{-i q_x n^- \lambda /2 } e^{-i \frac{\pi}{4}  \term{n^+ + (n^+)^2/2 + (n^-)^2/2}} \nonumber \\
&\qquad \times e^{-(x + \lambda / 2 -n^+ \lambda / 4)^2 - \term{q_y - n^- \lambda / 4}^2} \frac{\tilde \Delta'(q_x,q_y)}{\sqrt{v_F^2 q_z^2 + \vert{\tilde \Delta'(q_x,q_y)}\vert^2}}.
\end{align}
We apply the transformation $n \mapsto n+1 , n' \mapsto n' +1$, which effectively realizes $n^+ \mapsto n^+ +2$, $n^- \mapsto n^-$. We find, 
\begin{align}
\expect{\psi_L(x + \lambda/2, y+\pi / \lambda)\psi_R(x + \lambda / 2, y + \pi / \lambda)} &\propto \sum_{n,n'} \int_{q_x, q_y, q_z}  e^{i (n^+ + 2) (y +  \pi /\lambda)  \lambda/2} e^{-i q_x n^- \lambda /2 } e^{-i \frac{\pi}{4}(2n^+ + 4) }e^{-i \frac{\pi}{4}  \term{n^+ + (n^+)^2/2 + (n^-)^2/2}} \nonumber \\
&\qquad \times e^{-(x  -n^+ \lambda / 4)^2 - \term{q_y - n^- \lambda / 4}^2} \frac{\tilde \Delta'(q_x,q_y)}{\sqrt{v_F^2 q_z^2 + \vert{\tilde \Delta'(q_x,q_y)}\vert^2}}, \\
&= e^{iy\lambda}  \sum_{n,n'} e^{i \frac{\pi}{2}n^+}e^{-i \frac{\pi}{2}n^+} \int_{q_x, q_y, q_z}  e^{i n^+  y } e^{-i q_x n^- \lambda /2 } e^{-i \frac{\pi}{4}  \term{n^+ + (n^+)^2/2 + (n^-)^2/2}} \nonumber \\
&\qquad \times e^{-(x  -n^+ \lambda / 4)^2 - \term{q_y - n^- \lambda / 4}^2} \frac{\tilde \Delta'(q_x,q_y)}{\sqrt{v_F^2 q_z^2 + \vert{\tilde \Delta'(q_x,q_y)}\vert^2}}, \\ 
&\propto  e^{iy\lambda} \expect{\psi_L(x , y)\psi_R(x, y )}
\end{align}
Since, again, the proportionality constant is the same, we have $ \expect{\psi_L(x + \lambda/2, y+\pi / \lambda)\psi_R(x + \lambda / 2, y + \pi / \lambda)} =  e^{iy\lambda} \expect{\psi_L(x , y)\psi_R(x, y )}$. Therefore, we find that the zeros must form a lattice with translation vectors $(0, 2\pi l_B^2 / \lambda)$ and $(\lambda / 2, \pi l_B^2 / \lambda)$, which we call a triangular lattice.

\end{document}